\PassOptionsToPackage{dvipsnames}{xcolor}
\documentclass[a4paper,twocolumn,10pt,accepted=2026-11-16]{quantumarticle}
\pdfoutput=1

\usepackage[utf8]{inputenc}
\usepackage[T1]{fontenc}
\usepackage[english]{babel}
\usepackage{amsmath,amssymb,amsthm,mathtools,bm}
\usepackage{bbm,physics}
\usepackage{graphicx}
\usepackage{booktabs}
\usepackage{xspace}
\usepackage[numbers,sort&compress]{natbib}
\usepackage{hyperref}

\DeclareMathSymbol{\shortminus}{\mathbin}{AMSa}{"39}

\renewcommand\ket[1]{{|{#1}\rangle}}
\begin{document}

\title{Analytical and Compressed Simulation of Noisy Stabilizer Circuits}
\author{Paul Aigner}
\email{paul.aigner@uibk.ac.at}
\affiliation{Institut f\"ur Theoretische Physik, Universit\"at Innsbruck, Technikerstra{\ss}e 21a, 6020 Innsbruck, Austria}
\author{Jasmin Matti}
\affiliation{Institut f\"ur Theoretische Physik, Universit\"at Innsbruck, Technikerstra{\ss}e 21a, 6020 Innsbruck, Austria}
\author{Maria Flors Mor-Ruiz}
\affiliation{Institut f\"ur Theoretische Physik, Universit\"at Innsbruck, Technikerstra{\ss}e 21a, 6020 Innsbruck, Austria}
\author{Julius Wallnöfer}
\affiliation{Institut f\"ur Theoretische Physik, Universit\"at Innsbruck, Technikerstra{\ss}e 21a, 6020 Innsbruck, Austria}
\author{Wolfgang Dür}
\affiliation{Institut f\"ur Theoretische Physik, Universit\"at Innsbruck, Technikerstra{\ss}e 21a, 6020 Innsbruck, Austria}

\date{November 17, 2026}

\begin{abstract}
 We develop analytical and algorithmic techniques that enable efficient simulation of a broad class of noisy stabilizer circuits. We derive closed-form expressions of expectation values for tensor product of Paulis in circuits with non-deterministic Pauli measurements, enabling efficient exact evaluation of Pauli expectation values that avoids explicit density matrix construction and enables direct noise parameter sweeps. We introduce a circuit compression framework that can reduce the per-sample cost of weak simulation in general noisy stabilizer circuits, including deterministic measurements, by separating parameter-independent preprocessing from sampling. Finally, we extend the analytical framework beyond its standard domain to include a small number of deterministic measurements, general rotations, and non-diagonal noise channels. Our results provide a unified framework for both analytical and weak simulation of noisy stabilizer circuits and corresponds to an extension of the noisy stabilizer formalism introduced in \cite{PhysRevA.107.032424}. Applications range from calculation of the expectation values of entanglement witnesses, determination of reduced states, to energy evaluation.
\end{abstract}

\maketitle


\section{Introduction}
\label{Sec.Introduction}

Classical simulation of quantum circuits is a key tool for the verification and benchmarking of quantum devices~\cite{Villalonga2019qflex}, and for the analysis of fault-tolerant quantum computation~\cite{PhysRevA.70.052328,Gidney2021stimfaststabilizer}. For general noisy quantum dynamics, however, brute-force simulation quickly becomes intractable, since the explicit propagation of mixed states by density matrices incurs an exponential overhead in the number of qubits. A major exception is provided by the stabilizer setting: pure stabilizer states undergoing Clifford operations and Pauli measurements admit an efficient classical description and update rule, as captured by the Gottesman-Knill theorem and its algorithmic refinements~\cite{osti_319738,PhysRevA.70.052328,PhysRevA.73.022334,Gidney2021stimfaststabilizer}.

In practice, noise is unavoidable, and this immediately raises the question of how far the efficient stabilizer paradigm can be extended beyond the noiseless setting. Existing high-performance stabilizer simulators address this challenge primarily from a sampling perspective. In particular, modern simulators such as \textit{Stim}~\cite{Gidney2021stimfaststabilizer} combine tableau-based preprocessing with Pauli-frame propagation in order to generate large numbers of samples from noisy stabilizer circuits. This weak-simulation viewpoint is very successful when the objective is to estimate output statistics or detector events from Monte Carlo samples. Its cost per shot is low, but the precision of any estimated observable is fundamentally limited by sampling noise and improves only as $O(M^{-1/2})$ with the number of shots $M$.

In this work we pursue a complementary route. Rather than focusing exclusively on sampling, we develop methods for the analytical treatment of noisy stabilizer circuits. Our starting point is the noisy stabilizer formalism (NSF)~\cite{PhysRevA.107.032424}, introduced to efficiently describe Pauli-diagonal noise acting on graph-state protocols under Clifford operations and Pauli measurements. The original NSF already shows that, by separating the stabilizer-state update from the update of the noise operators, one can avoid explicit mixed-state propagation whenever the relevant structure of the circuit is preserved. Here we show that this viewpoint can be pushed substantially further.

Our first main contribution is an efficient analytical method for Pauli expectation values in a broad class of noisy stabilizer circuits, namely the fragment generated by Clifford gates, Pauli-diagonal noise channels, and non-deterministic Pauli measurements. For this class, we derive closed-form expressions for Pauli expectation values directly in terms of commutation relations between the observable and the noise operators. In contrast to Monte Carlo estimators, these expressions yield exact expectation values for the specified noise model, with no statistical fluctuations. Moreover, once the commutation structure has been determined, the result depends only parametrically on the noise strengths, so that sweeps over channel parameters can be performed without repeated re-simulation of the circuit.

Our second main contribution is a compression framework for noisy stabilizer circuits. The central idea is to separate parameter-independent preprocessing from parameter- or sample-dependent evaluation. By absorbing compatible Clifford operations and measurement steps into an updated stabilizer reference description, one obtains a reduced circuit representation that can lower the per-sample cost of subsequent weak simulation. In this sense, our approach does not compete with Pauli-frame sampling~\cite{PhysRevA.99.062337,Gidney2021stimfaststabilizer}; rather, it complements it. Sampling-based methods remain advantageous when only moderate precision is required, whereas our analytical treatment becomes attractive in the high-accuracy regime and in applications where one needs exact Pauli expectation values, repeated parameter scans, or access to families of observables derived from the same compressed description.

Beyond the non-deterministic measurement fragment, we further show how the analytical NSF viewpoint can be extended to deterministic Pauli measurements, general rotations, and non-diagonal channels. In these cases the simulation cost increases, and in particular deterministic measurements lead to a branching structure whose worst-case complexity grows exponentially in their number. Nevertheless, the resulting framework remains efficient for circuits containing only a limited number of such operations.

\begin{figure}
    \centering
    \includegraphics[width=1\linewidth]{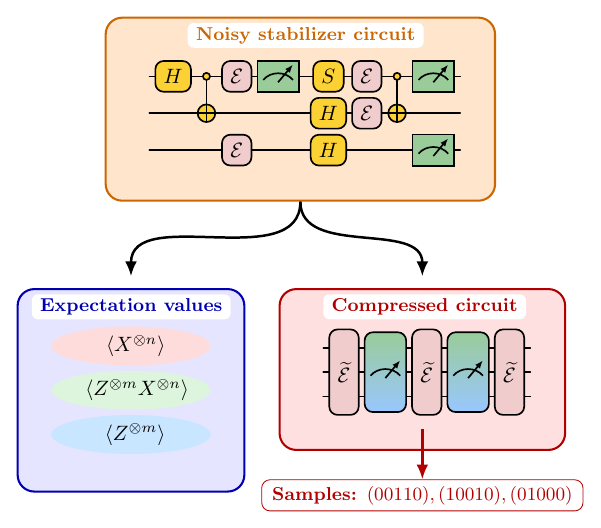}
    \caption{Overview schematic illustrating the formalism’s capabilities: noisy stabilizer circuits can be efficiently processed to compute arbitrary Pauli expectation values or to generate compressed circuit representations, enabling more efficient sampling.}
    \label{fig:1}
\end{figure}

The methods developed here are relevant for several tasks that naturally reduce to Pauli expectation values, including fidelity estimation, entanglement detection, calculation of reduced density matrices on small subsystems, Bell-inequality witnesses, and energy evaluation for Hamiltonians with polynomially many Pauli terms. The different working regimes of the formalism have been schematically illustrated in Fig.~\ref{fig:1}.

The remainder of the paper is organized as follows. In Sec.~\ref{Sec.Background} we review the stabilizer formalism and reformulate the NSF in a circuit-oriented language. In Sec.~\ref{sec:analytical_sim} we derive efficient analytical formulas for Pauli expectation values in the non-deterministic measurement fragment and discuss applications. In Sec.~\ref{Sec:noisy_stabilizer_circuit_compression} we introduce our compression procedure and explain how it reduces the cost of weak simulation. In Sec.~\ref{Sec:Generalizations} we describe extensions to deterministic measurements and more general channels. Finally, in Sec.~\ref{Sec:conclusion} we provide a summary of the results and an outlook for future work.

\section{Background}
Below, we briefly review the standard notation and key results for stabilizer states~\cite{nielsen2010quantum}, and present a reformulation of the NSF~\cite{PhysRevA.107.032424} tailored to our purposes.
\label{Sec.Background}
\subsection{Stabilizer formalism}

A (non-trivial) stabilizer group $\mathcal{S} < \mathcal{P}_n$ is an abelian subgroup of the $n$-qubit Pauli group 
\begin{equation}
    \mathcal{P}_n=\{\pm1, \pm i\}\times \{\mathbbm{1},X,Y,Z\}^{\otimes n},
\end{equation}
that does not contain $-\mathbbm{1}$~\cite{nielsen2010quantum}. That is, for all $S_i, S_j \in \mathcal{S}$,
\begin{equation}
    [S_i, S_j] = 0, \quad S_i \neq -\mathbbm{1}.
\end{equation}

The group is finitely generated as $\mathcal{S} = \langle g_1, \dots, g_r \rangle$, with $r \leq n$, where the generators are independent. In the case $r = n$, the stabilizer group uniquely specifies a stabilizer state $\ket{S}$ through
\begin{equation}
    S_l \ket{S} = \ket{S}, \quad \forall S_l \in \mathcal{S}.
\end{equation}
Thus, the state can be represented efficiently by its generators rather than by an exponentially large density matrix.

Stabilizer states can be efficiently simulated under Clifford operations and Pauli measurements, as formalized by the Gottesman--Knill theorem~\cite{osti_319738}. The central idea is to track the evolution of the stabilizer generators instead of the quantum state.

The Clifford group $\mathcal{C}_n$ is defined as the normalizer of the Pauli group,
\begin{equation}
    \mathcal{C}_n = \{U \in \mathcal{U}(2^n) : U P U^\dagger \in \mathcal{P}_n \ \forall P \in \mathcal{P}_n\},
\end{equation}
where $\mathcal{U}(2^n)$ denotes the group of all unitaries. 
Consequently, Clifford operations map Pauli operators to Pauli operators under conjugation. For a stabilizer state $\ket{S}$ and a Clifford unitary $C$, the transformed state $C\ket{S}$ is stabilized by
\begin{equation}
    \widetilde{\mathcal{S}} = \langle C g_1 C^\dagger, \dots, C g_n C^\dagger \rangle.
\end{equation}
Hence, Clifford gates can be simulated efficiently by updating the generators via conjugation.

Pauli measurements can also be treated efficiently. The projectors onto the $\pm 1$ eigenspaces of a Pauli operator $P$ are given by
\begin{equation}
    P_{P,\pm}=\frac{\mathbbm{1} \pm P}{2}.
\end{equation}
Two cases arise. If $\pm P \in \mathcal{S}$, the measurement outcome is deterministic and the stabilizer group remains unchanged. If $\pm P \notin \mathcal{S}$, then $P$ anticommutes with at least one generator $g_i$, which we choose without loss of generality to be $g_1$. One can choose a generating set such that only the single generator $g_1$ anticommutes with $P$, while all others commute, by multiplying $g_1$ to the original generators $g_i$ which anticommute with $P$~\cite{osti_319738,PhysRevA.70.052328}. The post-measurement stabilizer group is then
\begin{equation}
    \widetilde{\mathcal{S}} = \langle \pm P, \widetilde{g}_2, \dots, \widetilde{g}_n \rangle,
\end{equation}
where the sign is determined by the random measurement outcome.

\begin{figure}
    \centering
     \begin{minipage}{1\columnwidth}
    \centering
    \includegraphics[width=\linewidth]{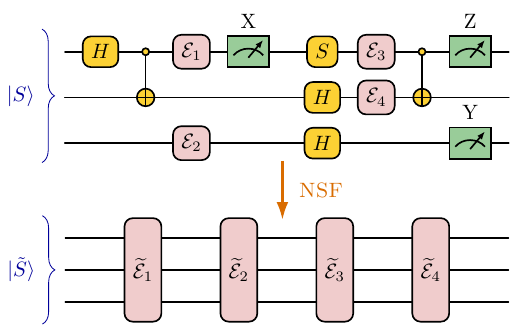}
    \vspace{0.3em}
    {\small (a)}
\end{minipage}
     \begin{minipage}{0.39\columnwidth}
    \centering
    \includegraphics[width=1\linewidth]{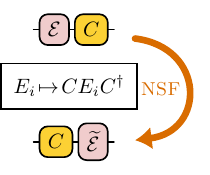}
    \vspace{0.3em}
    {\small (b)}
\end{minipage}
 \begin{minipage}{0.59\columnwidth}
    \centering
    \includegraphics[width=1\linewidth]{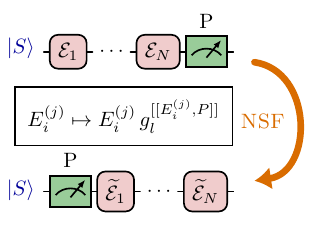}
    \vspace{0.3em}
    {\small (c)}
\end{minipage}
    \caption{Overview of the NSF.
\textbf{(a)}~Illustrates the framework: an initial stabilizer state is processed by a noisy stabilizer circuit composed of non-deterministic measurements, Clifford gates, and Pauli-diagonal noise channels. The NSF maps this circuit to an equivalent representation with an updated initial stabilizer state, where the Clifford operations and measurements are absorbed into the state, and the noise channels are updated while preserving their Pauli Kraus rank.
\textbf{(b)}~Shows the update rule for Clifford gates: each Pauli noise term $E_j$ is transformed by conjugation with the Clifford as the gate is pushed through the noise channel.
\textbf{(c)}~Update rule for non-deterministic Pauli measurements: each Pauli noise term $E_j^{(i)}$ that anti-commutes with the measurement observable is mapped to a commuting operator by multiplication with a stabilizer generator $g_l$ that also anti-commutes with the observable. The operator $[[A,B]]$ equals $0$ if $A$ and $B$ commute, and $1$ otherwise. }
    \label{fig:2}
\end{figure}

\subsection{Noisy stabilizer formalism (NSF)} \label{Subsec.Noisy stabilizer formalism}

The NSF was introduced in Ref.~\cite{PhysRevA.107.032424} as an efficient method to fully describe and analyze Pauli-diagonal noise acting on graph states manipulated by Clifford operations and Pauli measurements. It was subsequently applied to the study of graph-state extraction from noisy resource states in quantum networks~\cite{MorRuiz2025imperfectquantum,10479180} and to the analysis of a novel quantum repeater~\cite{morruiz2025mergingbasedquantumrepeater}. The formalism was later extended to higher-dimensional quantum systems~\cite{gqfw-x72s}. Its efficiency relies on the independent treatment and update of graph states and noise operators. The central mechanism enabling this separation is the set of commutation relations between Pauli noise operators and manipulation operators, referred to as \textit{update rules}. These rules allow one to apply manipulation operators directly to the graph state, that are efficiently described via the Gottesman-Knill theorem, while independently updating the noise channels.

In this work, we reformulate the original formalism to focus on noisy stabilizer circuits rather than graph state manipulation. Specifically, we consider a stabilizer state $\ketbra{S}$ acted on by Pauli-diagonal noise channels of the form
\begin{equation}
\mathcal{E}_i(\bullet)=\sum_j \lambda_j^{(i)} E_j^{(i)} (\bullet) E_j^{(i)},
\end{equation}
where $E_j^{(i)} \in \{\mathbbm{1},X,Y,Z\}^{\otimes n}$ is a tensor product of Pauli operators, such that
\begin{equation} \label{equ:nsf_standard}
\rho=\mathcal{E}_1 \cdots \mathcal{E}_N (\ketbra{S}).
\end{equation}
The Pauli-diagonal channels considered here cover several standard noise models. These include the single-qubit depolarizing channel
\begin{equation}
\mathcal{E}(\bullet)
=
(1-p)(\bullet)
+
\frac{p}{3}\bigl(X(\bullet) X + Y(\bullet) Y + Z(\bullet) Z\bigr),
\end{equation}
as well as multi-qubit channels such as the two-qubit depolarizing channel
\begin{equation}
\mathcal{E}(\bullet)
=
(1-p)(\bullet)
+
\frac{p}{15}\sum_{(i,j)\neq(0,0)} (\sigma_i \otimes \sigma_j)(\bullet)(\sigma_i \otimes \sigma_j),
\end{equation}
where $\{\sigma_i\}_{i=0}^3 = \{\mathbbm{1}, X, Y, Z\}$. In particular, this framework also includes correlated noise of the form
\begin{equation}
\mathcal{E}(\bullet)
=
(1-p)(\bullet)
+
p\,\sigma_i^{\otimes m}(\bullet) \sigma_i^{\otimes m}.
\end{equation}
We refer to the expression in Eq.~\eqref{equ:nsf_standard} as the \textit{NSF standard form}. Our objective is to maintain this efficient representation under Clifford operations, Pauli measurements, and the inclusion of additional Pauli-diagonal noise channels, see Fig.~\ref{fig:2}~(a).

Under Clifford operations, each noise channel is updated by conjugating its individual noise operators $E_i$ with the Clifford unitary, independently of the state. Simultaneously, the stabilizer state is updated according to the stabilizer formalism, see Fig.~\ref{fig:2}~(b).

For Pauli measurements, we distinguish between two cases: \textit{deterministic} and \textit{non-deterministic} measurements, depending on whether the measurement outcome is fixed or random on the pure state. The NSF introduced in \cite{PhysRevA.107.032424} applies only to the non-deterministic case, as these are the relevant transformations for pure graph states. The restriction to non-deterministic measurements becomes clear from the following considerations.

Let $P$ denote a non-deterministic Pauli observable. By definition, $\pm P \notin \mathcal{S}$, where $\mathcal{S}$ is the stabilizer group of $\ket{S}$, implying that $P$ anticommutes with at least one stabilizer generator $g_l$. Starting from the NSF standard form (Eq.~\eqref{equ:nsf_standard}), 
we expand the final noise channel:
\begin{equation}
\rho= \sum_i \lambda_i^{(N)} \mathcal{E}_1\cdots\mathcal{E}_{N-1}(E_i^{(N)}\ketbra{S}E_i^{(N)}).
\end{equation}
We then insert the stabilizer generator $g_l$ into each term such that the modified noise operators commute with $P$:
\begin{equation}
\begin{aligned}
\rho&=\sum_i \lambda_i^{(N)}
  \mathcal{E}_1\cdots\mathcal{E}_{N-1}\Bigl(
\\
&\qquad
  E_i^{(N)}
  g_l^{[[E_i^{(N)},P]]}
  \ketbra{S}
  g_l^{[[E_i^{(N)},P]]}
  E_i^{(N)}
  \Bigr)
\\
&\equiv
  \mathcal{E}_1\cdots\mathcal{E}_{N-1}
  \widetilde{\mathcal{E}}_N(\ketbra{S}),
\end{aligned}
\end{equation}
where $[[A,B]] = 0$ if $A$ and $B$ commute, and $1$ otherwise. The channel $\widetilde{\mathcal{E}}_N$ denotes the updated noise channel. Since Pauli-diagonal noise channels commute, this procedure can be applied independently to each channel, yielding
\begin{equation}
\rho=\widetilde{\mathcal{E}}_{1} \cdots \widetilde{\mathcal{E}}_N (\ketbra{S}),
\end{equation}
where all updated channels now commute with $P$. Consequently, the measurement can be commuted through the noise and applied directly to the stabilizer state, which can then be efficiently updated, see Fig.~\ref{fig:2}~(c).

In contrast, for deterministic measurements, the observable $\pm P$ is contained in the stabilizer group. In this case, the noise channels cannot be made to commute with the measurement, and the above procedure fails. Therefore, deterministic measurements require leaving the NSF standard form; this will be addressed in Sec.~\ref{Sec:Analytical_noisy_circuit_simulation}.

Additional Pauli-diagonal (circuit-level) noise can be incorporated by appending new noise channels to the existing noise channel string of the NSF standard form:
\begin{equation}
    \mathcal{E}_{N+1}(\rho)=\mathcal{E}_{1}\cdots \mathcal{E}_{N+1} (\ketbra{S}).
\end{equation}

Noisy Clifford gates and Pauli measurements can be modeled by inserting a noise channel before and/or after the operation. For Clifford gates, it is always sufficient to place the noise channel only before the gate, since any noise after the gate can be conjugated through the Clifford and absorbed into the preceding noise channel. For measurements, the situation depends on whether the measured qubits are discarded. If they are discarded after readout, it is sufficient to include only a noise channel before the measurement. If the qubits are reused later, as in mid-circuit measurements, noise channels after the measurement must also be taken into account.

Overall, this methodology allows one to efficiently reduce the non-deterministic fragment of noisy stabilizer circuits to the NSF standard form. The associated computational overhead is analyzed in Appendix~\ref{app:NSF_complexity}.

In many applications, the object of interest is only a reduced subsystem. This occurs, for example, in Clifford measurement-based quantum computing (MBQC)~\cite{PhysRevLett.86.5188}, where single-qubit Pauli measurements on a stabilizer resource state implement a Clifford operation on the remaining unmeasured qubits, and in graph-state manipulation~\cite{Freund_2025,Szymanski2026usefulentanglement,basak2024improvedroutingmultipartyentanglement,Sen2025multipartite,10.1098/rsta.2017.0325}, where local Clifford operations and local Pauli measurements transform a resource graph state into one tailored to a particular application. If the stabilizer state factorizes with respect to a qubit $v$, i.e. $\ket{S}=\ket{\Psi}_v \otimes \ket{S'}$, one can trace out this qubit by removing its local Pauli components from all noise operators:
\begin{equation}
\mathrm{tr}_v(\rho)=\widetilde{\mathcal{E}}_1 \cdots \widetilde{\mathcal{E}}_N \ketbra{S'},
\end{equation}
where each $\widetilde{\mathcal{E}}_j$ is the reduced noise channel obtained by removing the Pauli support on $v$. The factorization condition can be checked efficiently, for example via the stabilizer–graph state equivalence and inspection of the corresponding adjacency matrix~\cite{PhysRevA.69.022316}.

If the resulting reduced stabilizer state is small, applying the updated noise channels using the density matrix formalism remains efficient, enabling the computation of the final mixed state without large density matrices. This constitutes the primary application domain of the original NSF. However, when the final state is large, explicit application of the noise channels becomes inefficient. In such cases, one must instead work directly with the NSF standard form to extract relevant properties of the mixed state while preserving computational efficiency. This constitutes the main focus of the present work.
\subsubsection{Deterministic versus non-deterministic measurements}

Here, we highlight the key differences between deterministic and non-deterministic Pauli measurements in noisy stabilizer circuits and connect them to the NSF.

A first key difference is their information content. For non-deterministic measurements, the outcomes remain uniformly distributed over $\pm 1$, even in the presence of Pauli-diagonal noise. As a result, their measurement statistics carry no information about the noise. In contrast, deterministic measurement outcomes are distributed according to the probability that the measured observable has been flipped by the noise channel prior to measurement. Therefore, deterministic Pauli measurements directly probe properties of the noise acting on the stabilizer state.

A second difference concerns the post-measurement noise channel. For deterministic measurements, the outcome reveals whether the measured observable was flipped, and the measurement therefore projects the noise channel onto either the flip-free branch or the flip-error branch. For non-deterministic measurements, the noise channel can be chosen to commute with the measurement projector, so its form is independent of the measurement outcome. In particular, the post-measurement states associated with different non-deterministic outcomes differ only by an outcome-dependent Pauli operator, while the noise channel itself remains unchanged. Hence, for non-deterministic measurements all outcome branches can be dealt with at the same time largely without extra effort, while for deterministic measurements only one branch at a time is considered. 
\subsection{Stabilizer tableau simulation and Pauli frame sampling} \label{subsec:stabilizer_tableau_pauli_frame}
Here, we review how state-of-the-art stabilizer simulators, such as \textit{Stim}~\cite{Gidney2021stimfaststabilizer} and related generalizations~\cite{haenel2026tsimfastuniversalsimulator,brandl2026quickquditsframeworkefficientsimulation,kabir2025sdimquditstabilizersimulator, fang2026lightstimframeworkqecprotocol,chase2026clifftfastexactsimulation}, efficiently generate samples from noisy stabilizer circuits.
\subsubsection{Stabilizer tableau}
As a first step, we introduce the concept of a stabilizer tableau, which is utilized to efficiently store and update a noiseless stabilizer state, under Pauli measurements and Clifford gates. 

A stabilizer tableau simulator represents an $n$-qubit stabilizer state using a binary data structure that tracks a generating set of the stabilizer group under Clifford evolution. The simulator maintains a tableau of size $O(n^2)$ over the finite field $\mathbb{F}_2$, which enables efficient simulation of Clifford circuits including intermediate measurements.

In a standard formulation, c.f. \cite{PhysRevA.70.052328}, the tableau is a $2n \times (2n+1)$ binary matrix whose rows correspond to Pauli generators. The first $n$ rows are typically interpreted as destabilizers and the remaining $n$ rows as stabilizers. The destabilizer group is a list of commuting Pauli operators $\{d_1,\dots, d_n\}$, which anti-commute with exactly one stabilizer generator, i.e. $[[d_i,g_j]]=\delta_{ij}$. The columns are partitioned into $X$-components, $Z$-components, and a phase bit. Explicitly, the tableau can be written as
\begin{equation}
T = [\, X \mid Z \mid r \,],
\end{equation}
where $X, Z \in \mathbb{F}_2^{2n \times n}$ and $r \in \mathbb{F}_2^{2n}$ encodes the sign of each generator. Each row $i$ of the tableau specifies a Pauli operator of the form
\[
P_i = (-1)^{r_i} \bigotimes_{j=1}^n i^{X_{i,j} Z_{i,j}} X^{X_{i,j}} Z^{Z_{i,j}},
\]
so that the full tableau compactly represents a set of $2n$ independent Pauli operators generating the Pauli group.

Clifford gates act on the tableau via linear transformations over $\mathbb{F}_2$ that preserve the underlying symplectic structure. These transformations can be implemented using bit-wise operations on the tableau rows. 

Pauli measurements are handled by checking commutation relations between the measured observable and the current stabilizer generators; if anti-commutation occurs, the tableau is updated through row operations analogous to Gaussian elimination, ensuring that the post-measurement state is correctly represented.
If the observable lies in the stabilizer group (up to a sign), the measurement outcome is fixed and can be inferred without modifying the underlying quantum state.

\subsubsection{Pauli frame sampling}
Pauli-frame simulation \cite{PhysRevA.99.062337} is a technique to efficiently generate large numbers of samples from noisy
stabilizer circuits without re-running a full tableau simulation for every shot.
The central idea is to separate the simulation into two components:
(1)~a noiseless reference trajectory and (2)~classically tracked
Pauli errors (the Pauli frame). Note that the Pauli frame technique is independent of the exact implementation or inner workings of a stabilizer simulator.
\subsubsection{Pauli Frames}
A Pauli frame is a classical record specifying, for each qubit,
whether the current state differs from a reference state by an $X$ and/or $Z$
operator.  Each qubit therefore carries two bits $(x_i, z_i)$.
Clifford gates update the frame by conjugation, which acts as a deterministic linear transformation on the bits $(x_i, z_i)$.
This update is cheaper than evolving a full stabilizer tableau.
\subsubsection{Noise Accumulation}
For Pauli channels, one does not
apply noise directly to the stabilizer state.
Instead, at each noise location a Pauli operator is sampled from the channel
distribution and multiplied into the Pauli frame.

\subsubsection{Reference Sample and Outcome Generation}
A Pauli-frame simulator yields only the parity flips
that should be applied to measurement outcomes, where the measurements are restricted without loss of generality to Pauli-$Z$ measurements. 
Therefore, one first computes a noiseless reference shot
using a stabilizer tableau simulator.
For each subsequently sampled Pauli frame, the measurement outcome is
\begin{equation}
m = m_{\mathrm{ref}} \oplus f ,
\end{equation}
where $f$ is the frame bit indicating the presence of an $X$ or $Y$ error on
the measured qubit.
This reduces sampling cost to tracking frame updates, enabling Monte-Carlo simulation of large stabilizer circuits.

\section{Analytical Simulation of the Non-Deterministic Measurement Fragment} \label{sec:analytical_sim}

We now present our approach for efficiently computing Pauli expectation values, including expectation values of tensor products of local Pauli operators, using the NSF formalism for noisy stabilizer circuits. Our approach is restricted to circuits composed of Clifford gates and non-deterministic Pauli measurements, which ensures tractability while still capturing a broad and practically relevant class of quantum processes under noise.

Prominent examples of such circuits include Clifford MBQC, where computation proceeds via adaptive sequences of Pauli measurements on entangled resource states, as well as protocols for measurement-driven manipulation of resource stabilizer states. These settings highlight the applicability of our method to distributed and noise-affected quantum information processing scenarios.

\subsection{Efficient analytical calculation of Pauli expectation values}
\label{sec.:effpauliexp}
\begin{figure}
    \centering
    \includegraphics[width=.8\linewidth]{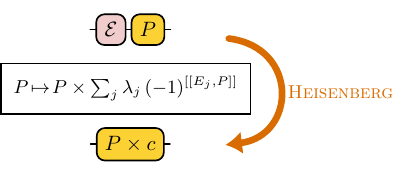}
    \caption{Illustration of the Heisenberg action of a Pauli diagonal channel $\mathcal{E} (\bullet) = \sum_j \lambda_j E_j (\bullet) E_j$ on a Pauli observable $P$.}
    \label{fig:3}
\end{figure}

We show that Pauli expectation values 
\begin{equation}
    \langle \bigotimes_{i=1}^n \sigma_{j_i}^{(i)} \rangle_\rho=\textrm{tr}(\bigotimes_{i=1}^n \sigma_{j_i}^{(i)} \rho)
\end{equation}
of the non-deterministic fragment of noisy stabilizer circuits can be computed analytically and efficiently. Any noisy stabilizer circuit in the non-deterministic fragment can be written in NSF standard form as
\begin{equation}
\rho=\mathcal{E}_1 \cdots \mathcal{E}_N (\ketbra{S}),
\end{equation}
see Sec.~\ref{Subsec.Noisy stabilizer formalism}. Thus, this mixed-state representation provides an efficient description of the output of any noisy stabilizer circuit restricted to non-deterministic measurements, including noisy Clifford MBQC circuits and graph-state manipulation protocols.

The key idea is to work in the Heisenberg picture. Rather than applying Pauli-diagonal noise channels to the state, we apply their Heisenberg action to the Pauli observable. For Pauli-diagonal channels, each Pauli operator is preserved up to a scalar factor. In particular, for a stabilizer element \( S_l \in \mathcal{S} \),
\begin{equation} \label{equ:analytical_exp_value}
\begin{aligned}
\langle S_l\rangle_\rho
&= \mathrm{tr}\!\left(S_l \, \mathcal{E}_1 \cdots \mathcal{E}_N \ketbra{S}\right) \\
&= \prod_{i=1}^{N} \sum_j \lambda^{(i)}_{j} \, (-1)^{[[E_j^{(i)}, S_l]]},
\end{aligned}
\end{equation}
which follows from the Heisenberg action of Pauli-diagonal channels,
\begin{equation}
\mathcal{E}_i^T(S_l) = \sum_j \lambda^{(i)}_{j} \, (-1)^{[[E_j^{(i)}, S_l]]} \, S_l,
\end{equation}
depicted in Fig.~\ref{fig:3}, together with the stabilizer property $S_l \ketbra{S} = \ketbra{S}$.
For any Pauli operator $\pm P \notin \mathcal{S}$, the expectation value vanishes, since $P$ maps the stabilizer state to an orthogonal state, yielding zero trace.

Let $K_i$ denote the number of Pauli terms with nonzero weight in the noise channel $\mathcal{E}_i$. Evaluating Eq.~\eqref{equ:analytical_exp_value} then requires $\sum_{i=1}^N K_i$ commutation checks between the noise operators and $S_l$.
Each check determines the parity of the number of anticommuting local Pauli pairs on the common support.
Since each local check and parity update takes constant time and the common support contains at most $n$ qubits, the worst-case time complexity is $\mathcal{O}(n\sum_{i=1}^N K_i)$.
Defining $K_{\max}=\max_i K_i$ gives the bound $\mathcal{O}(nNK_{\max})$.
For fixed $K_{\max}$, the cost is linear in each of $n$ and $N$.

Since the expectation value in Eq.~\eqref{equ:analytical_exp_value} is analytic, one can derive a general expression for fixed noise operators $ E_j^{(i)} $ and variable noise strengths $ \lambda_j^{(i)}$. The required commutation relations are independent of the specific values of $ \lambda_j^{(i)}$, so the result depends only parametrically on these quantities. This yields an analytical form of the stabilizer expectation value in terms of $ \lambda_j^{(i)} $, enabling efficient sweeps over noise parameters.

\begin{table*}[t]
\centering
\caption{Quality metrics accessible from analytical Pauli expectation
values. Exact evaluation is efficient for polynomially many required
Pauli terms.
See Appendix~\ref{Appendix:survey_of_quality_metrics} for details.}
\label{tab:quality_metrics}
\small
\setlength{\tabcolsep}{4pt}
\renewcommand{\arraystretch}{1.1}
\begin{tabular}{ccc}
\hline\hline
\textbf{Metric} & \textbf{Property} & \textbf{Evaluation} \\
\hline
\shortstack{Entanglement witnesses\\
\cite{PhysRevA.72.022340,li2025detectinggenuinemultipartiteentanglement}} &
\shortstack{Entanglement\\certification} &
\shortstack{Polynomially many\\stabilizer correlators.} \\[0.3em]

\shortstack{Reduced states and local entropies\\
\cite{PhysRevA.92.012305,Ruskai02}} &
\shortstack{Mixedness and\\correlations} &
\shortstack{Stabilizers supported\\on small subsystems.} \\[0.3em]

\shortstack{Bell inequalities\\
\cite{PhysRevLett.95.120405,PhysRevLett.124.020402}} &
\shortstack{Nonlocality\\certification} &
\shortstack{Polynomially many\\Pauli correlators.} \\[0.3em]

\shortstack{Hamiltonian (energy)\\
\cite{Sun2025stabilizerground}} &
\shortstack{Energy\\benchmarking} &
\shortstack{Polynomially many Pauli terms\\
(e.g., fixed-$k$-body interactions).} \\[0.3em]

\shortstack{Average MBQC fidelity\\
\cite{2603.13753}} &
\shortstack{Computational\\performance} &
\shortstack{Weighted stabilizer sum;\\
efficient estimation by sampling terms.} \\
\hline\hline
\end{tabular}
\end{table*}

Table~\ref{tab:quality_metrics} summarizes quality metrics for noisy
stabilizer states, covering a broad range of physically relevant
properties. These metrics can be evaluated
or estimated efficiently from analytical Pauli expectation values.
Definitions and further details are provided in
Appendix~\ref{Appendix:survey_of_quality_metrics}.

Pauli expectation values in the non-deterministic noisy stabilizer-circuit fragment can be estimated by sampling or computed analytically.
Sampling-based simulators such as Stim~\cite{Gidney2021stimfaststabilizer} offer low per-shot cost, with a worst-case sample complexity of $\mathcal{O}(\epsilon^{-2})$ for additive error $\epsilon$ at fixed confidence.
Our analytical method evaluates the Heisenberg action of Pauli-diagonal noise channels directly, without sampling error.
Sampling can be preferable for modest accuracy, whereas analytical evaluation can be advantageous at high accuracy or for noise-parameter sweeps.
The relative runtime depends on the circuit, observable, and implementation. In Appendix~\ref{app:sampling_vs_analytical_exp_values}, we provide error bounds and a cost comparison.

\subsection{Example} \label{subsec:example}
Here, we utilize the techniques described above, to determine an $n$-partite entanglement witnesses~\cite{PhysRevA.72.022340} of a noisy graph state~\cite{PhysRevA.69.062311, Hein2006GraphStates}, which are obtained from different noisy resource graph states via a sequence of Pauli measurements and Clifford operations. Specifically, we consider the generation of a noisy MBQC or fusion based computation~\cite{Bartolucci2023} resource state, obtained by the fusion of caterpillar states~\cite{PhysRevA.111.052604}. For the calculation of the updated noise maps, we utilize the \textit{noisy-graph-states} package \cite{Wallnoefer2025NoisyGraphStates}, which is a graph state based implementation of the original NSF formulation. 

A graph state is a special type of stabilizer states, where the canonical generators of the stabilizer have the form $g_j=X_j \prod_{i \in N(j)}Z_i$, where $N(j)$ is the neighborhood of a vertex labeled by $j$ of a graph $G=(V,E)$. We can map graph states to other graph states, by a certain class of Clifford operations as well as Pauli measurements with Clifford corrections~\cite{PhysRevA.69.062311, Hein2006GraphStates}. 
\subsubsection{Fusion of noisy caterpillar states}
\begin{figure}
    \centering
\includegraphics[width=1\linewidth]{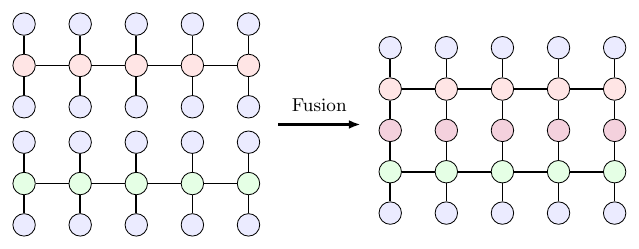}
    \caption{Illustration of the fusion of two caterpillar graph states.}
    \label{fig:5}
\end{figure}

We consider the fusion of caterpillar graph states, which is a linear cluster graph state with a certain number of leaves at each linear cluster vertex. We consider caterpillars with two leaves at each linear cluster vertex, and when fusing two caterpillar graph states, we fuse always one leaf per node, see Fig.~\ref{fig:5} for an illustration. Fusion or merging operations on a graph state can be described by a Clifford gate plus a projective Pauli measurement~\cite{Pirker_2018, PhysRevA.74.052316}
\begin{equation}
    \mathrm{Fuse}_{s,t} (\bullet) =\mathrm{tr}_t(P_{Z_t,\pm} CNOT_{s \rightarrow t} (\bullet)),
\end{equation}
up to an outcome dependent Pauli correction. This setting is motivated by photonic platforms, where entangling gates are difficult~\cite{PhysRevA.111.052604}. There a standard approach is to have a quantum emitter with a spin degree of freedom~\cite{Lindner2009,Thomas2022}, which can generate caterpillar type graph states, and then to utilize type-II fusion~\cite{PhysRevLett.95.010501,Lee2023graphtheoretical,Lobl2024losstolerant} to generate more complicated graph state structures~\cite{PhysRevA.111.052604}. 

Figure~\ref{fig:6}~(a) shows the graph-state $m$-partite entanglement witness expectation value~\cite{PhysRevA.72.022340}
\begin{equation}
    \langle W_G \rangle = (m-1)-\sum_{v\in V(G)} \langle g_v\rangle,
\end{equation}
for noisy resource states of size $m=(2k+1)n$ obtained by fusing $k$ noisy caterpillar graph states, with $k\in\{2,4,8\}$. We model the noise by local depolarizing noise applied initially to the caterpillar graph states with a depolarizing probability $p=0.001$. The expectation value of the witness is plotted as a function of the caterpillar length $n$.

Negative values of $\langle W_G\rangle$ certify genuine multipartite entanglement, while non-negative values indicate that this witness no longer detects entanglement. Each caterpillar consists of a linear backbone of length $n$, with two spike qubits attached to every backbone vertex. Adjacent caterpillars are fused using the simple merging operation $\mathrm{Fuse}_{s,t}$, producing a larger noisy resource state.

As $n$ increases, noise suppresses the stabilizer expectation values and thus increases the witness expectation value. The different curves illustrate how this behavior depends on the number $k$ of fused caterpillars. Solid lines indicate the regime $\langle W_G\rangle<0$, whereas dotted lines mark the regime in which the witness becomes non-negative.
\subsubsection{Noisy generation and fusion of caterpillar states with imperfect operations}
The analysis above, demonstrates the effect of initial noise being transformed by ideal operations. Now we consider the case where the operations are noisy, as well as analyze the noisy generation process of the noisy caterpillar graph states. We model noisy one qubit gates by an depolarizing channel prior to the application of the gate, whereas we model noisy two-qubit gates by applying two-qubit depolarizing channels prior to the gates. The details on the graph state generation protocol, can be found in Appendix~\ref{app:graph_state_generation}.

\begin{figure}
    \centering
     \begin{minipage}{1\columnwidth}
    \centering
\includegraphics[width=\linewidth]{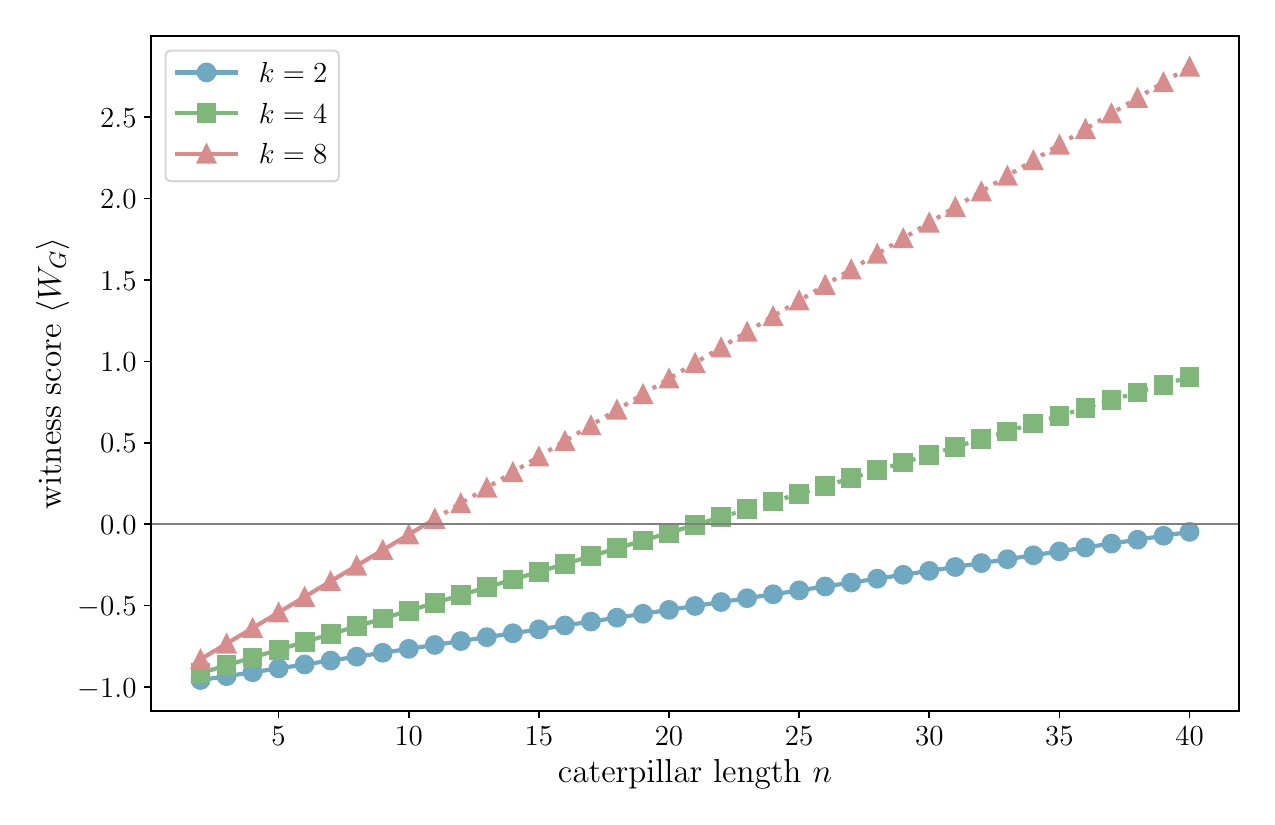}
    \vspace{0.3em}
    {\small (a)}
\end{minipage}
\begin{minipage}{1\columnwidth}
    \centering
\includegraphics[width=\linewidth]{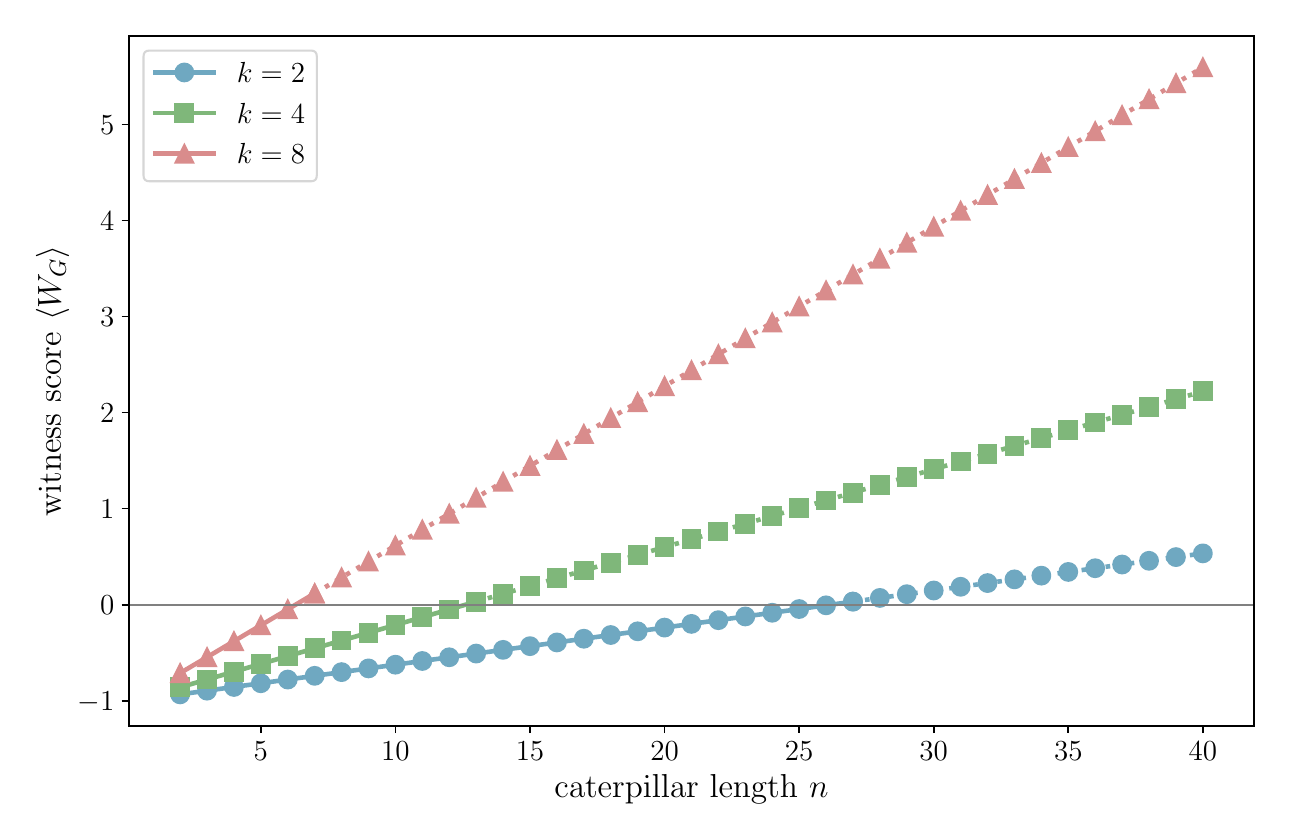}
    \vspace{0.3em}
    {\small (b)}
\end{minipage}

    \caption{\textbf{(a)} $m(n,k)=(2k+1)n$-partite entanglement witness score $\langle W_G\rangle$ for an noisy MBQC resource state of size $m(n,k)$ obtained by fusing $k$ caterpillar states subject to local depolarizing noise with depolarizing probability $p=0.001$, each of length $n$ with two leaves attached to every backbone node, where $W_G = m - 1 - \sum_j g_j$. \textbf{(b)} $m(n,k)$-partite entanglement witness score $\langle W_G\rangle$ for a noisy MBQC resource state of size $m(n,k)$, obtained by first generating $k$ caterpillar states according to the protocol described in Appendix~\ref{app:graph_state_generation}, each of length $n$ and with two leaves attached to every backbone node, and then fusing them with noisy operations. Single-qubit operation noise is modeled by a depolarizing channel with probability $p_1=0.0005$, and two-qubit operation noise by a two-qubit depolarizing channel with probability $p_2=0.001$, with both channels applied before the corresponding operation.}
    \label{fig:6}
\end{figure}

For single-qubit operations we choose a depolarizing probability of $p_1=0.0005$, and for two-qubit operations $p_2=0.001$. We then determine the $m$-partite entanglement witness for different system sizes by varying the backbone length $n$ and the number of caterpillar states $k$; see Fig.~\ref{fig:6}~(b). As in the case of initial noise only, the witness score increases with system size. For the chosen noise parameters, however, the score from the noisy generation process exceeds the entanglement-witness threshold already for smaller systems.

\section{Noisy stabilizer circuit compression}
\label{Sec:noisy_stabilizer_circuit_compression}
\begin{figure}
    \centering
    \includegraphics[width=1\linewidth]{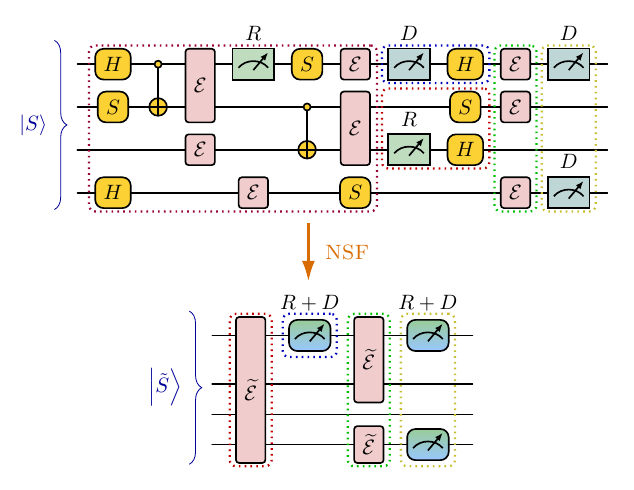}
    \caption{Illustration of circuit compression. A noisy stabilizer circuit with noise channels $\mathcal{E}$, Clifford gates, and random ($R$) and deterministic ($D$) Pauli measurements acting on $\ketbra{S}$ is converted into an equivalent compressed circuit with updated noise channels $\widetilde{\mathcal{E}}$ and combined measurement blocks ($R+D$) acting on $\ketbra{\tilde{S}}$. }
    \label{fig:4}
\end{figure}
We describe how the NSF framework can be used to compress general noisy stabilizer circuits. The goal is to construct an equivalent representation with fewer effective operations, thereby potentially reducing the per-sample simulation cost while shifting complexity into a preprocessing stage, see Fig.~\ref{fig:4} for an illustration of the methodology.

Let $\mathcal{C}$ be a circuit composed of Clifford unitaries, Pauli measurements, and Pauli-diagonal noise channels. We partition the Pauli measurements into
deterministic measurements $M_i^{(D)}$, whose outcome is fixed on the \textit{pure reference stabilizer state}, which will be defined below, and random measurements $M_i^{(R)}$, whose outcomes are $\pm1$ with equal probability.
The compression framework below, in contrast to the analytical framework in Sec.~\ref{sec.:effpauliexp}, is not restricted to the non-deterministic measurement fragment. As deterministic measurements inform on the noise properties of a noisy stabilizer state, they are required for quantum error correction~\cite{Roffe03072019}, as well as for entanglement purification~\cite{Dur_2007}. Hence, the compression framework can also be applied to accelerate sampling of logical error probabilities.

\subsection{Methodology}
We split the noisy stabilizer circuit into a part prior to the first deterministic  measurement and the part after the deterministic measurement. The noisy stabilizer circuit prior to the first deterministic measurement belongs to the non-deterministic fragment of noisy stabilizer circuits. Hence the NSF procedure yields the NSF standard form
\begin{equation}
\rho = \mathcal{E}_1 \cdots \mathcal{E}_N \ketbra{S},
\end{equation}
and the first deterministic measurement is in general kept explicitly as an operator,
\begin{equation}
\rho' =
M_1^{(D)}
\left(
\mathcal{E}_1 \cdots \mathcal{E}_N
\ketbra{S}
\right)
M_1^{(D)}.
\end{equation}
If the deterministic projector does not commute with the noise channels, it cannot be absorbed into the stabilizer state as in the standard NSF framework; hence if it does not commute we explicitly keep it.
Subsequent Clifford operations act by conjugation, while measurements and noise channels are either propagated or appended depending on commutation relations. This structure: noise channels interleaved by a string of measurement projectors acting on a stabilizer state, serves as the template for the fully general case.

 At any stage, the state admits a representation of the form
\begin{equation}
\begin{aligned}
\rho &=
\prod_l \mathcal{M}_l^{(R)}\mathcal{M}_l^{(D)}
\prod_j \mathcal{E}_j\cdots
\\
&\quad \cdots
\prod_i \mathcal{M}_i^{(R)}\mathcal{M}_i^{(D)}
\prod_k \mathcal{E}_k \ketbra{S},
\end{aligned}
\end{equation}
where $\mathcal{M}_i^{(D/R)} (\bullet) = M_i^{(D/R)} (\bullet) M_i^{(D/R)}$ and we call $\ketbra{S}$ the \textit{reference stabilizer state} with respect to which the measurement type of a Pauli measurement is defined. This structure is preserved under all circuit updates, as becomes clear from the following processing rules.

\paragraph{Noise channels.}
When inserting a new Pauli-diagonal noise channel $\mathcal{E}$ we simply append it to the description.

\paragraph{Deterministic measurements.}
A measurement $\mathcal{M}_k^{(D)}=\mathcal{M}_{P_k,s_k}$ has deterministic outcome $s_k=\pm1$ on the reference stabilizer state $\ket{S}$ if
$P_k=s_k\prod_i g_i^{u_i}$, where $g_i$ are its stabilizer generators and $u_i\in\{0,1\}$.
The measurement map is commuted through retained measurements and Pauli noise channels according to
\begin{equation}
\begin{aligned}
    \mathcal{M}_k^{(D)}\circ\mathcal{M}_{Q,r}
    &=\mathcal{M}_{Q,r}\circ\mathcal{M}_k^{(D)},
    &&[P_k,Q]=0,\\
    \mathcal{M}_k^{(D)}\circ\mathcal{E}
    &=\mathcal{E}\circ\mathcal{M}_k^{(D)},
    &&[P_k,E_j]=0\ \forall E_j,
\end{aligned}
\end{equation}
where $E_j$ are the Pauli noise terms of $\mathcal{E}$.
If it reaches the reference state, it is absorbed using
$\mathcal{M}_k^{(D)}(\ketbra{S})=\ketbra{S}$.
If it encounters an identical retained measurement, the two maps merge by idempotence,
$(\mathcal{M}_k^{(D)})^2=\mathcal{M}_k^{(D)}$.
Otherwise, it is retained explicitly.

\paragraph{Clifford unitaries.}
For a Clifford unitary $U$, define $\mathcal{U}(X)=UXU^\dagger$.
Propagate it through retained projectors and Pauli channels by conjugating
their Pauli operators:
\begin{align}
    \mathcal{U}\circ\mathcal{M}_{P,s}
    &=\mathcal{M}_{UPU^\dagger,s}\circ\mathcal{U},\\
    \mathcal{U}\circ\mathcal{E}
    &=\tilde{\mathcal{E}}\circ\mathcal{U},
\end{align}
with $
    \tilde{\mathcal{E}}(\bullet)
    =\sum_j p_j(UE_jU^\dagger)(\bullet)(UE_jU^\dagger)^\dagger.$
On the pure reference state, we utilize the standard stabilizer formalism to update the state.
These updates preserve the present measurement-noise block structure.

\paragraph{Random measurements.}
Analogously to the standard NSF we aim to absorb random measurements in the reference stabilizer state, which results in a compressed circuit. In order for a random measurement to be absorbable into the reference stabilizer state, the necessary condition for the random measurement is to commute with all preceding measurement blocks.

Assuming, the random measurement commutes with all preceding measurement blocks, we have to make the noise channels commuting with the random measurement. In order to do that we aim to insert stabilizers in the error channels, which anti-commute with the observable associated to the random measurement, analogous to the standard NSF, see Sec.~\ref{Subsec.Noisy stabilizer formalism}. At the same time, we require that the inserted stabilizers commute with the measurement operator blocks, as non-commutativity would lead to potentially different outcomes in different error channel branches, leaving the efficient regime. Trivially, the deterministic measurements commute with the inserted stabilizers, as their observables are by definition contained in the stabilizer of the reference stabilizer state. A priori the random measurements in the measurement blocks anti-commute with some stabilizer generators, hence they do not trivially commute with the stabilizer generator we want to insert for the modification of the error channels. To the aim of constructing a suitable stabilizer generator to insert, we utilize the concept of the destabilizer group generated by $\{d_1,\dots,d_n\}$, see Sec.~\ref{subsec:stabilizer_tableau_pauli_frame}. As the random Pauli observables $P_j^{(R)}$ associated to the random Pauli measurements are not contained in the stabilizer group, we know $ P_j^{(R)}=\pm d_1^{\alpha_1^j}\cdots d_n^{\alpha_n^j}g_1^{\beta_1^j}\cdots g_n^{\beta_n^j}$, such that $\exists \alpha_i^{j} \neq 0$. At the same time we can write any potential adjustment stabilizer from the stabilizer group as $S_l=g_1^{t_1}\cdots g_n^{t_n}$. Hence, in order to make the error channels commute with the observable $P_0^{(R)}$, while simultaneously leaving the measurements associated to observables $\{P_j^{R}\}_{j=1}^{m_R}$ invariant, we require \begin{equation} \label{equ:cond_stabilizer_existence}
    \begin{aligned}
        \langle \boldsymbol{\alpha}^0 , \boldsymbol{t} \rangle&=1 \mod 2, \\
        \langle \boldsymbol{\alpha}^j , \boldsymbol{t} \rangle&=0 \mod 2, \quad j \in \{1,\dots,m_R\}.
    \end{aligned}
\end{equation}
Here, $\langle \boldsymbol{v},\boldsymbol{w}\rangle=\sum_{i=1}^n v_iw_i \pmod 2$ denotes the standard inner product for binary vectors $\boldsymbol{v}=(v_1,\dots,v_n)$ and $\boldsymbol{w}=(w_1,\dots,w_n)$ in $\mathbb{F}_2^n$. Thus, iff $\boldsymbol{\alpha}^0 \not \in \textrm{span}_{\mathbbm{F}_2}\{\boldsymbol{\alpha}^j\}_{j=1}^{m_R}$ there exists a $\boldsymbol{t}\in \mathbbm{F}_2^n$ satisfying condition Eq.~\eqref{equ:cond_stabilizer_existence}. One, can determine $\boldsymbol{t}$ by Gaussian elimination over $\mathbbm{F}_2$ by building a system of equations according to Eq.~\eqref{equ:cond_stabilizer_existence}. Therefore, we can absorb a random measurement in the reference stabilizer iff 
\begin{enumerate}
    \item The random measurement observable $P_0$ commutes with all measurement blocks.
    \item The random measurement observable $ P_0 = \pm d_1^{\alpha_1^0}\cdots d_n^{\alpha_n^0}g_1^{\beta_1^0}\cdots g_n^{\beta_n^0}$ and the random measurement observables contained in the measurement blocks $ P_j = \pm d_1^{\alpha_1^j}\cdots d_n^{\alpha_n^j}g_1^{\beta_1^j}\cdots g_n^{\beta_n^j}$ fulfill Eq.~\eqref{equ:cond_stabilizer_existence}, for some $\boldsymbol{t} \in \mathbbm{F}_2^n$. 
\end{enumerate}
If those conditions are not fulfilled the random measurement cannot be absorbed and has to be appended. 
    
These rules preserve the alternating measurement–noise block structure throughout the circuit traversal, yielding the final compressed representation in the same form. This representation absorbs Clifford operations and compatible measurements into the stabilizer description, while retaining a compressed sequence of measurements and noise channels.

For an $n$-qubit circuit, let $C$ and $M$ denote the numbers of elementary one- or two-qubit Clifford gates and Pauli measurements, respectively, let $R\le M$ count measurements processed as random relative to the current reference state, and let $K=\sum_i K_i$ be the total number of explicitly listed Pauli noise terms.
Using binary Pauli strings and an initially supplied stabilizer-destabilizer tableau, a straightforward implementation has the conservative worst-case operation count
\begin{equation}\label{eq:compression_complexity}
\begin{aligned}
\mathcal{O}\!\Bigl(
  n^2&+nK+C(n+M+K)
\\
&+Mn(M+K)+n^2M(R+1)
\Bigr).
\end{aligned}
\end{equation}
Thus, constructing the compressed representation is polynomial in the explicitly represented circuit and noise data. Appendix~\ref{app:compression_complexity} gives a more detailed account of the individual update costs.

\subsection{Sampling-speed up for the non-deterministic measurement fragment}
Here, we come back to the case of the non-deterministic measurement fragment. However, instead of using the fact that we can efficiently compute Pauli expectation values, we harness the effectively reduced circuit depth that can speed up sampling of the circuit. As we are in the non-deterministic fragment the compressed circuit is given by the NSF standard form
\begin{equation}
\rho=\mathcal{E}_1\cdots\mathcal{E}_N (\ketbra{S}).
\end{equation}
Therefore, given an initial circuit build up out of $m_R$ non-deterministic measurements, $m_C$ Clifford gates, $N$ error channels, is compressed to a circuit consisting only $N$ updated error channels.  

This compressed circuit can then be used for accelerated sampling of some properties, like the state fidelity, of the output a Clifford MBQC or graph state extraction protocol, utilizing the Pauli frame simulation framework~\cite{PhysRevA.99.062337, Gidney2021stimfaststabilizer} with potentially reduced complexity per sampling shot.

\section{Generalizations} \label{Sec:Generalizations}
Below, we introduce extensions of the analytical simulation framework beyond the non-deterministic measurement fragment of noisy stabilizer circuits. Specifically, we generalize the framework to include a finite amount of deterministic measurements as well as non-Pauli-diagonal channels. 
\subsection{Analytical noisy stabilizer circuit simulation} \label{Sec:Analytical_noisy_circuit_simulation}

Here, we demonstrate how, by using the NSF description of stabilizer states together with the ability to compute Pauli expectation values, one can analytically track Pauli measurements and Clifford operations on stabilizer states in the presence of Pauli-diagonal noise channels. The computational complexity of this analytical tracking is polynomial in the system size, as well as in the number of Clifford operations and non-deterministic measurements, but becomes exponential in the number of deterministic measurements.

\subsubsection{Methodology}

Deterministic measurements occur when the measurement observable is contained in the stabilizer group, i.e., $\pm P \equiv S_l \in \mathcal{S}$. The post-measurement state can then be written as
\begin{equation}
    \rho' \propto \underbrace{\frac{\mathbbm{1}\pm S_l}{2}}_{\equiv P_{S_l,\pm}}
    \underbrace{\mathcal{E}_1\cdots\mathcal{E}_N(\ketbra{S})}_{\equiv \rho}
    \frac{\mathbbm{1} \pm S_l}{2}.
\end{equation}
Note that $\rho'$ is only proportional to the action of the projector, since normalization by the probability $\langle P_{S_l,\pm} \rangle_\rho$ is required. This quantity can be computed efficiently using the results of Sec.~\ref{sec.:effpauliexp}.

A direct computation shows that the measurement projector $P_{S_l,\pm}$ commutes with the noisy state $\rho$. Using the idempotency of projectors, this simplifies to
\begin{equation}
    \rho' \propto P_{S_l,\pm} \rho.
\end{equation}
Expanding the projector yields
\begin{equation} \label{equ:det_meas_proj}
    \rho' \propto \frac{\rho}{2} \pm \frac{S_l \rho}{2}.
\end{equation}

We now analyze the term $S_l \rho$:
\begin{equation}
    \begin{aligned}
        S_l \rho 
        &= S_l \mathcal{E}_1 \cdots \mathcal{E}_N \ketbra{S} \\
        &= S_l \sum_{j} \lambda_{j}^{(1)} E_j^{(1)} \mathcal{E}_2 \cdots \mathcal{E}_N \ketbra{S} E_j^{(1)} \\
        &= \sum_{j} (-1)^{[[E_j^{(1)}, S_l]]} \lambda_{j}^{(1)} E_j^{(1)} S_l \mathcal{E}_2 \cdots \mathcal{E}_N \ketbra{S} E_j^{(1)} \\
        &\equiv \widetilde{\mathcal{E}}_1 S_l \mathcal{E}_2 \cdots \mathcal{E}_N \ketbra{S} \\
        &= \widetilde{\mathcal{E}}_1 \cdots \widetilde{\mathcal{E}}_N \ketbra{S},
    \end{aligned}
\end{equation}
where $\widetilde{\mathcal{E}}_i$ denote updated maps. These maps have the same operators and the same absolute values of the Kraus coefficients as $\mathcal{E}_i$, but may differ by a sign in front of the coefficients. As a result, they are generally non-physical channels. However, since they remain diagonal in the Pauli basis, their treatment within the NSF framework is mathematically equivalent, as only the operators (and not the weights) are updated.

This leads to the following expression for the updated noisy state:
\begin{equation}
    \rho' = \frac{1}{2 \langle P_{S_l,\pm} \rangle_\rho}
    \left(
        \mathcal{E}_1 \cdots \mathcal{E}_N \ketbra{S}
        \pm
        \widetilde{\mathcal{E}}_1 \cdots \widetilde{\mathcal{E}}_N \ketbra{S}
    \right).
\end{equation}

Thus, after a single deterministic measurement, the number of terms that must be tracked doubles. Iterating this procedure leads, in the worst case, to $2^{m}$ terms, where $m$ is the number of deterministic measurements. Consequently, for $\log(n)$ deterministic measurements, the description remains efficient.

Importantly, standard NSF updates arising from Clifford gates and non-deterministic measurements can always be applied to a single term, since the noise operators are identical across all deterministically updated channels, up to a sign. Expectation values can then be computed term by term using linearity of the trace and summed at the end. Hence, any stabilizer expectation value can be computed with polynomial overhead in the number of noise operators, Clifford operators, and non-deterministic measurements, while having an exponential overhead in the number of deterministic measurements.   

After $m$ deterministic measurements, a fixed outcome branch contains at most $2^m$ algebraic terms. There are at most $2^m$ deterministic-measurement outcome strings, so explicitly tracking all such branches requires at most $4^m$ terms. This remains polynomial for $m=\mathcal O(\log n)$. Tracking only selected outcome branches, as relevant in many application such as entanglement purification, can reduce the cost.

\subsection{Including general channels}
\label{sec:gen-rot}
Here, we show how to extend the analytical simulation of noisy stabilizer states, to arbitrary rotations or noise channels; enabling analytical computation of Pauli expectation values also in the presence of finite number of non-stabilizer operations. The extension of the NSF to general channels has been demonstrated in the standard formulation of the NSF~\cite{matti2025extensions}; and hence is restricted to the non-deterministic fragment. Here, we show how to treat general channels in the new formulation, how to include deterministic measurements, as well as how to determine analytical Pauli expectation values.

We can write an arbitrary quantum channel $\mathcal{E}$ as a linear combination of Pauli matrices
\begin{equation} \label{equ:general_channel}
\mathcal{E} (\bullet) = \sum_{P,Q \in \mathcal{P}_n  } \alpha_{P,Q} P (\bullet)  Q.
\end{equation}
Note, that the channel description in Eq.~\eqref{equ:general_channel} also includes general rotations and measurements. Hence given, the NSF standard form of a noisy stabilizer state $\rho=\mathcal{E}_1\cdots \mathcal{E}_N(\ketbra{S})$, we can expand the action of an arbitrary channel in the Pauli basis terms
\begin{equation}
    \mathcal{E} (\rho) = \sum_{P,Q \in \mathcal{P}_n } \alpha_{P,Q}  P \rho Q.
\end{equation}
Then, we can update the noise channels for each linear term
\begin{equation} \label{equ:unit_action}
    \mathcal{E}(\rho)=\sum_{P,Q \in \mathcal{P}_n} \alpha_{P,Q} \mathcal{E}_1^{P,Q} \dots \mathcal{E}_N^{P,Q} P \ketbra{S} Q,
\end{equation}
where in each term the updated non-physical diagonal noise channels $\mathcal{E}^{P,Q}$ are obtained from the commutation relations of the noise channel with the channel $P (\bullet) Q$.

As a next step, we show how to update noisy stabilizer states which where subject to arbitrary channels
\begin{equation}
    \rho=\sum_{P,Q \in R \subset \mathcal{P}_n} \alpha_{P,Q}\mathcal{E}_{1}^{P,Q}\cdots \mathcal{E}_{N}^{P,Q} P \ketbra{S}Q.
\end{equation}

Clifford operations, can be performed via conjugating each noise channel independently. The channel $P (\bullet) Q $ is updated in the same fashion, and the pure state $\ketbra{S}$ can be updated via the pure state stabilizer formalism. 

For Pauli measurements, we differentiate between measurements which are deterministic on the reference stabilizer state $\ketbra{S}$ and those which are non-deterministic. 

For non-deterministic Pauli measurements associated to observable $A$ we determine the action of the projector:
\begin{equation}
    P_{A,\pm} \sum_{P,Q \in R \subset \mathcal{P}_n} \alpha_{P,Q} \mathcal{E}_{1}^{P,Q}\cdots \mathcal{E}_{N}^{P,Q} P \ketbra{S}Q P_{A,\pm}.
\end{equation}
As the projector is a linear map, we look at the action on a general term
\begin{equation}
P_{A,\pm}\mathcal{E}_{1}^{P,Q}\cdots \mathcal{E}_{N}^{P,Q} P \ketbra{S}Q P_{A,\pm},
\end{equation}
where we neglect the coefficients for convenience. As in the standard NSF, we apply each noise channel $\mathcal{E}_{i}^{P,Q}$ individually to the pure state term $P \ketbra{S}Q$, and make the noise channel commute via insertion of suitable stabilizer. The only difference is that due to the non-diagonal Pauli channel $P (\bullet) Q$ the stabilizers will enter with potentially flipped signs in the noise channel terms, depending whether on the commutation relation of the required stabilizer with $P$ and $Q$. Then, we obtain a updated noise channel string which commutes with the measurement projector
\begin{equation}
\tilde{\mathcal{E}}_{1}^{P,Q}\cdots \tilde{\mathcal{E}}_{N}^{P,Q} P_{A,\pm}  P \ketbra{S}Q P_{A,\pm}
\end{equation}
Note the channels are not physical channels as the signs of the weights can be negative. Next, we concentrate on the term 
\begin{equation}
   P_{A,\pm}^{(v)}  P \ketbra{S}Q P_{A,\pm}^{(v)}.
\end{equation}
We commute the projectors through the channel $P (\bullet) Q$ yielding
\begin{equation}
    P P_{A,\pm(-1)^{[[A,P]]}}\ketbra{S}P_{A,\pm(-1)^{[[A,Q]]}} Q.
\end{equation}
Then, we apply the projectors onto the state, using the pure state stabilizer formalism, yielding
\begin{equation}
    P J(\pm(-1)^{[[A,P]]})\ketbra{S'}J(\pm(-1)^{[[A,Q]]}) Q, 
\end{equation}
where $J$ is an outcome dependent correction Pauli gate, which fixes the state $\ket{S'}$. Now redefining $P J(\pm(-1)^{[[A,P]]})$ as $\tilde{P}$ and $J(\pm(-1)^{[[A,Q]]}) Q$ as $\tilde{Q}$ yields the original form before the measurement. Normalization, i.e. calculation of the outcome probability, can be performed via calculating the expectation value of the Pauli observable using the methodology from Sec.~\ref{sec.:effpauliexp}. Explicitly, we calculate
\begin{equation}
    \mathrm{tr}(P_{A,\pm} \rho).
\end{equation}
Using the linearity of the trace, and the normalization of the state, we only have to compute terms of the form
\begin{equation}
    \mathrm{tr}(A \mathcal{E}_1 \cdots \mathcal{E
    }_N P \ketbra{S}Q)=q\mathrm{tr}(QAP \ketbra{S})=\tilde{q},
\end{equation}
where one obtains $q$ from the Heisenberg action of the noise channels on the measurement observable $A$, and $\tilde{q}=\pm q \delta_{ \pm QAP \in S}$, with $\delta_{X \in Y}=1$ if $X \in Y$ and otherwise $\delta_{X \in Y}=0$. After all terms have been added with their corresponding coefficients, the associated projector probability can be computed efficiently in the number of terms, enabling proper normalization. Hence for the non-deterministic measurement fragment, with $\log(n)$ number of general channels, with bounded number of Pauli terms, the number of efficiently trackable terms stays polynomial and we can efficiently track the evolution of the state. As for deterministic measurements, the number of different measurement branches grows exponentially, hence for $\log(n)$ number of measurements, tracking of all possible measurement outcome trajectories stays efficient. Note that the exponential number of measurement paths is not unique to the noisy tracking of the state, but also existing for the pure state version \cite{PhysRevA.70.052328}. In Appendix~\ref{app:special_non_diag_channels} we show that for special classes of noise channels and mixed input states the dependence on the random measurement outcomes can be contained to the pure-state representative, while the updates of the noise channels are outcome independent.
\subsubsection{Deterministic measurements}
For deterministic measurements with respect to the reference stabilizer state, we run into the same issue as in the standard NSF framework. We cannot make the measurement operator commuting with the noise channels. Hence, if we propagate the measurement to the state, we will expand the measurement operator analogous to the deterministic measurement in noisy stabilizer circuits, see Sec.~\ref{Sec:Analytical_noisy_circuit_simulation}. We calculate the action of the deterministic Pauli measurement associated to the observable $B$:
\begin{equation}
    P_{B,\pm} \sum_{P,Q \in R \subset \mathcal{P}_n} \alpha_{P,Q} \mathcal{E}_{1}^{P,Q}\cdots \mathcal{E}_{N}^{P,Q} P \ketbra{S}Q P_{B,\pm}.
\end{equation}
As for the non-deterministic case we look at the action on a general term
\begin{equation}
\eta_{P,Q}\equiv P_{B,\pm}\mathcal{E}_{1}^{P,Q}\cdots \mathcal{E}_{N}^{P,Q} P \ketbra{S}Q P_{B,\pm}.
\end{equation}
We can pull out the non-diagonal channel $P (\bullet) Q$, by updating the non-physical noise channels as well as the measurement projectors:
\begin{equation}
    \eta_{P,Q} 
    =P P_{B,c_P} \underbrace{\widetilde{\mathcal{E}}_{1}^{P,Q}\cdots \widetilde{\mathcal{E}}_{N}^{P,Q}  (\ketbra{S}) }_{\equiv \chi_{P,Q}}P_{B,c_Q} Q,
\end{equation}
where $c_P=\pm(-1)^{[[B,P]]}$ and $c_Q=\pm(-1)^{[[B,Q]]}$.
As, in the deterministic measurement case for noisy stabilizer circuits, see Sec.~\ref{Sec:Analytical_noisy_circuit_simulation}, the measurement projectors commute with $\chi_{P,Q}$ and we obtain
\begin{equation}
    \eta_{P,Q}=P P_{B,\pm(-1)^{[[B,P]]} } P_{B,\pm(-1)^{[[B,Q]]}} \chi_{P,Q} Q
\end{equation}
Hence, we see that the term $\eta_{P,Q}$ vanishes if the measurement observable $B$ has different commutation relations with $P$ and with $Q$. This yields
\begin{equation}
    \eta_{P,Q}=\delta([[B,P]]\oplus[[B,Q]])P P_{B,\pm (-1)^{[[B,P]]}}\chi_{P,Q} Q,
\end{equation}
with $\delta(x)=1$ for $x=0$ and $\delta(x)=0$ for $x\neq 0$, and $\oplus$ denoting addition modulo two. Now, the term $P_{B,\pm(-1)^{[[B,P]]}}\chi_{P,Q}$ is formally the same expression as in the deterministic case for noisy stabilizer circuits, see Eq.~\eqref{equ:det_meas_proj}, and we can perform the same expansion into two terms
\begin{equation}
P_{B,\pm(-1)^{[[B,P]]}}\chi_{P,Q}=\frac{1}{2}(\chi_{P,Q}\pm(-1)^{[[B,P]]}\chi'_{P,Q}),
\end{equation}
where $\chi'_{P_Q}=\widetilde{\widetilde{\mathcal{E}}}_{1}^{P,Q} \cdots \widetilde{\widetilde{\mathcal{E}}}_{N}^{P,Q}  \ketbra{S}$ is the operator $\chi_{P,Q}$ with updated non-physical error channels obtained through the commutation with the observable $B$. Therefore, by pulling back the channel $P (\bullet) Q$, we obtain for each non-vanishing term $\eta_{P,Q}$ two terms associated to the channel $P (\bullet) Q$. Normalization can be performed analogously to the non-deterministic measurement case.

Hence, we can describe a general quantum circuit with overhead exponential only in the number of non-diagonal noise channels, general rotations, and deterministic measurements relative to the stabilizer-state representative, while treating Clifford operations and non-deterministic measurements efficiently. This enables efficient estimation of Pauli expectation values for circuits containing few such costly operations.

\section{Conclusion and Outlook}\label{Sec:conclusion}

We have developed a framework for the simulation of noisy stabilizer circuits that combines two complementary perspectives: exact analytical evaluation of observables and compressed representations for accelerated sampling. The central observation is that, utilizing the noisy stabilizer formalism, one can avoid explicit propagation of the full mixed state and instead work with an efficiently updatable description consisting of a stabilizer reference state together with transformed noise operators. This makes it possible to access physically relevant quantities directly, without resorting to density-matrix methods.

For the fragment of noisy stabilizer circuits generated by Clifford gates, Pauli-diagonal noise, and non-deterministic Pauli measurements, we derived closed-form expressions for Pauli expectation values. This yields an efficient analytical method for a broad class of observables and diagnostic quantities, including reduced states on small subsystems, entanglement witnesses, Bell inequalities, and energies of Hamiltonians with polynomially many Pauli terms. In contrast to Monte Carlo approaches, the resulting expressions are exact for the specified noise model, free of statistical fluctuations, and naturally suited for parameter sweeps, since the relevant commutation data can be reused across different noise strengths.

At the same time, we showed that the same formalism also supports a compression viewpoint. By absorbing compatible Clifford operations and measurements into the reference stabilizer description, one obtains reduced noisy circuit representations for general noisy stabilizer circuits, that can lower the per-sample cost of subsequent weak simulation. In this sense, this approach complements the state-of-the-art sampling methods such as tableau-based Pauli-frame simulation.

We further extended the framework beyond the standard NSF setting by including deterministic measurements, general rotations, and non-diagonal channels. These extensions enlarge the scope of the method, but also show its limitations. In particular, deterministic measurements induce a branching structure whose cost grows exponentially in their number, so that full analytical tracking remains efficient only when such operations are sufficiently rare. This identifies a natural boundary between the efficiently tractable regime and the regime where one must either accept exponential overhead or revert to approximate or sampling-based methods.

Several directions for future work are promising. Since the noisy stabilizer formalism also extends to higher-dimensional systems~\cite{gqfw-x72s}, the results presented here could be generalized to qudits. It would also be interesting to apply the methods introduced here to noisy MBQC and to graph-state processing protocols with large outputs, such as multipartite GHZ-state generation~\cite{PhysRevA.100.052333} or the extraction of arbitrary graph states from resource graph states~\cite{Freund_2025}.

Note added: During the preparation of this manuscript, a related study~\cite{umbrarescu2026syqmamemoryefficientsymbolicexact} was published. That work independently develops an analytic framework for noisy stabilizer circuits with non-Clifford elements, which is focused on the analysis of quantum error correction. Our work was completed separately, utilizes different methods, and was not influenced by that publication.
A key distinction is that their non-Clifford elements are restricted to Pauli rotations. By contrast, our method can, in principle, support arbitrary gates and channels expressible with a bounded number of Pauli terms, including non-unital channels such as amplitude damping.   

\section*{Author contributions}
ChatGPT was used for assistance with language editing, manuscript organization and literature searches. All the ideas and derivations in this work are due to the authors, who take full responsibility for
its content.
\section*{Acknowledgments} We would like to thank Alena Romanova for their valuable feedback on the initial draft of this manuscript. This research was funded in whole or in part by the Austrian Science Fund (FWF) 10.55776/P36009, 10.55776/P36010, 10.55776/PAT1710825 and 10.55776/COE1. We also acknowledge support by the Austrian Research Promotion Agency
(FFG) under Contract Number 914030 (Next Generation EU). For open access purposes, the author has applied a CC BY public copyright license to any author accepted manuscript version arising from this submission. Finanziert von der Europ\"aischen Union - NextGenerationEU.

\bibliographystyle{quantum}
\bibliography{ref.bib}
\renewcommand\appendixname{Appendix}
\appendix

\section{Numerical complexity of the NSF} \label{app:NSF_complexity}

We analyze the numerical complexity of updating noise channels within the NSF framework. Since different noise channels can be updated independently, these updates can be parallelized. Therefore, it suffices to analyze the complexity for a single noise channel and sum the contributions afterward.

In an NSF update of a noise channel, the weights of the Kraus operators remain unchanged; only the operators themselves are updated. Furthermore, because the channels are Pauli-diagonal, the phases of the Pauli operators can be omitted. Consequently, it is sufficient to store only the non-zero-weight Pauli strings together with their associated weights.

Below, the tableau formulation of the NSF follows a recent implementation of the NSF~\cite{qcomm_uibk_noisy_stabilizers_d_2026}. We represent a noise channel using a stabilizer-tableau like structure
\begin{equation}
\mathbf{N} = [X|Z],
\end{equation}
called the \textit{noise tableau}. Here, $\mathbf{N}$ is a $K \times 2n$ binary matrix whose rows store the $K$ Pauli noise operators
\begin{equation}
P_i \cong (x^{(i)}_1,\dots,x^{(i)}_n \mid z^{(i)}_1,\dots,z^{(i)}_n)
\equiv \boldsymbol{n}_i \in \mathbb{F}_2^{2n}
\end{equation}
of the channel $\mathcal{N}$. This tableau omits phase information and does not require the operators to be independent, nor does it require $K=n$.

Clifford transformations act linearly on the noise tableau via the corresponding symplectic matrix. Assuming the standard generating set $\langle S, H, \text{CNOT} \rangle = \mathcal{C}_n$, each noise vector $\boldsymbol{n}_i$ can be updated in constant time $\mathcal{O}(1)$, since each gate modifies only a fixed number of bits. Thus, updating a single noise tableau requires $\mathcal{O}(K)$ operations. Summing over all noise channels, the total overhead per Clifford gate is
\begin{equation}
C_{\text{Clifford}}=\mathcal{O}\!\left(\sum_i K_i\right).
\end{equation}

For non-deterministic measurements, we restrict without loss of generality to $Z_r$ measurements. In this case, we first identify a stabilizer generator of the pure state that anti-commutes with $Z_r$. The pure-state stabilizer generators are stored in a tableau, and checking commutation with $Z_r$ requires constant time per generator. In the worst case, all $n$ generators must be checked, yielding complexity $\mathcal{O}(n)$. Let the first anti-commuting generator be $S_i \cong \boldsymbol{v}_i$. This procedure is standard in stabilizer tableau updates and can therefore be reused here.

Next, we check whether each noise vector $\boldsymbol{n}_j$ commutes with $Z_r$. If it commutes, no update is required. If it anti-commutes, we update the operator via
\begin{equation}
S_i P_j \cong \boldsymbol{v}_i \oplus \boldsymbol{n}_j ,
\end{equation}
which requires $\mathcal{O}(n)$ bit operations. In the worst case, this leads to a cost of $\mathcal{O}(Kn+n)$ for a single noise channel.

Importantly, the search for the anti-commuting stabilizer generator in the pure-state tableau is performed only once and can be reused for all noise channel updates. Consequently, updating all noise channels after a measurement requires
\begin{equation}
C_{\text{Meas}}=\mathcal{O}\!\left(n + n\sum_i K_i\right).
\end{equation}
\subsection{Reducing number of terms in a noise channel} \label{subsec:reducing_noise_terms}
It can be useful in some instances to truncate the number of terms in a noise channel, as it influences the complexity of operations with it. One, can reduce the number of terms in a Pauli channel if some Pauli noise operators are equivalent up to a stabilizer of the stabilizer group corresponding to the stabilizer state. This can be efficiently pairwise checked, via multiplying two Pauli noise operators $P=P_1P_2$, and checking whether the product is in the stabilizer. The question $P \in S$ can be efficiently checked, via computing commutators $[P,S_k]$, with the stabilizer generators $S_k$. If all commutators are zero then the product is in the stabilizer, and they are equivalent, and can be merged. If not, i.e. $\exists S_j$ such that $[P,S_j] \neq 0$, the Pauli noise operators are nonequivalent and cannot be merged.  
\subsection{Merging equivalent channels}

Reducing the number of noise channels, in addition to simplifying individual channels, lowers the overall tracking complexity. A natural approach is to merge channels via multiplication. However, this generally increases the number of Pauli Kraus terms in the resulting channel, and thus the complexity.

The Pauli Kraus rank remains unchanged if the support of one channel is invariant under multiplication with the other, up to stabilizer equivalence. Formally, consider two noise channels $\mathcal{E}_1$ and $\mathcal{E}_2$ with supports
\begin{equation}
S_1 = \{E_j^{(1)} : \lambda_j^{(1)} > 0 \}, \quad
S_2 = \{E_j^{(2)} : \lambda_j^{(2)} > 0\}.
\end{equation}
Their product $\mathcal{E}_1 \mathcal{E}_2$ has non-increasing Pauli Kraus rank if
\begin{equation}
S_1 / \mathcal{S} = S_1 S_2 / \mathcal{S}
\quad \text{or} \quad
S_2 / \mathcal{S} = S_1 S_2 / \mathcal{S},
\end{equation}
where
\begin{equation}
S_1 S_2 = \{E_j^{(1)} E_k^{(2)} : \lambda_j^{(1)}  \lambda_k^{(2)} > 0 \}.
\end{equation}

In practice, this condition can be checked efficiently by constructing the support sets and verifying that all product operators lie within one of the original supports up to stabilizer equivalence, using the pairwise procedure described in Sec.~\ref{subsec:reducing_noise_terms}.
\subsection{Tracing out isolated vertices}
Often, one is interested only in a subsystem of a stabilizer state. This is particularly relevant when the pure state is separable with respect to qubit $v$, i.e., $\ket{S}=\ket{S'}\otimes\ket{\Psi}_v$. To test whether the state factorizes with respect to qubit $v$, one checks the commutation relations between the single-qubit Pauli operators $\{Z_v, X_v, Y_v\}$ and the stabilizer generators. If at least one of these operators commutes with all stabilizer generators, then the stabilizer state factorizes with respect to \(v\). If all of them anti-commute with at least one generator, then the state does not factorize. Since each commutation check requires only $\mathcal{O}(1)$ computational effort, the full procedure has worst-case complexity $\mathcal{O}(n)$.

If the separability criterion holds, the partial trace can be computed efficiently. In that case, there exists a Pauli operator $ P \in \{\pm Z_v, \pm X_v, \pm Y_v\}$ such that $P \in \mathcal{S}$. We first identify a generator $g_j$ with support on $v$, which requires at most $\mathcal{O}(n)$ single-bit commutation checks and therefore incurs a cost of $\mathcal{O}(n)$. Next, we multiply $g_j$ into every other generator supported on $v$, producing an updated set of generators $g_i'$ in which only $g_j$ acts on $v$. In the worst case, this step costs $\mathcal{O}(n^2)$. The partial trace is then obtained by removing $g_j$ from the generator list and deleting all local Pauli components of noise operators supported on qubit $v$. This final step has complexity $\mathcal{O}\!\left(n + \sum_i K_i\right)$.

Tracing out part of a stabilizer system reduces the number of degrees of freedom that must be tracked. It can therefore be computationally advantageous to discard separable subsystems as early as possible. Moreover, tracing out qubits can render distinct channels or noise terms equivalent, enabling further truncation as described above.

\section{Survey of computable quality metrics for noisy stabilizer states} \label{Appendix:survey_of_quality_metrics}

We survey a set of quality metrics for noisy stabilizer states that can be efficiently computed from Pauli expectation values. Given a noisy stabilizer state in NSF standard form,
\begin{equation}
    \rho = \mathcal{E}_1 \cdots \mathcal{E}_N (\ketbra{S}),
\end{equation}
which may arise from a noisy stabilizer circuit, all quantities considered below reduce to evaluating expectation values of Pauli operators. Within the NSF framework, these can be computed analytically and efficiently, making the following metrics particularly suitable for large-scale noisy stabilizer systems. The quantities below serve both as theoretical diagnostic tools for assessing noisy stabilizer states and as experimentally relevant metrics, thereby also representing simulations of practical diagnostic procedures.

Overall, these metrics demonstrate that a wide range of physically relevant properties can be reduced to the efficient evaluation of Pauli expectation values. This includes entanglement, fidelity, energies, non-locality, and computational utility. 

\subsection{Entanglement witnesses}
Entanglement witnesses~\cite{GUHNE20091} provide operational criteria to detect non-separability. For stabilizer and graph states, witnesses can be constructed directly from stabilizer generators. For example, bipartite and multipartite entanglement witnesses can be constructed as
\begin{equation}
    W = \mathbbm{1} - \sum_i c_i g_i,
\end{equation}
where $g_i$ are stabilizer generators and $c_i \in \mathbbm{R}$. If
\begin{equation}
    \mathrm{tr}(W\rho) < 0,
\end{equation}
then $\rho$ is certified to be entangled~\cite{PhysRevA.72.022340}. More refined witnesses for $k$-separability involve linear combinations of single- and two-body stabilizer generator correlators~\cite{li2025detectinggenuinemultipartiteentanglement}~. Since all such witnesses are linear combinations of $\textrm{poly}(n)$ Pauli operators, their expectation values can be evaluated efficiently, enabling scalable detection of multipartite entanglement in noisy stabilizer states.

\subsection{Reduced density matrices and local entropies}
Any reduced density matrix on a subsystem $A$ can be reconstructed by restricting the stabilizer expansion to operators supported on $A$,
\begin{equation}
    \rho_A = \frac{1}{2^{|A|}} \sum_{S_l \in \mathcal{S}:\,\mathrm{supp}(S_l)\subseteq A} \langle S_l \rangle S_l.
\end{equation}
Hence, for sufficiently small $|A|$, $\rho_A$ can be reconstructed
analytically and efficiently from $2^{|A|}$ Pauli expectation values
using the methods of Sec.~\ref{sec.:effpauliexp}.
Pure stabilizer states are uniquely specified by their reduced states on the supports of the stabilizer generators~\cite{PhysRevA.92.012305,PhysRevLett.134.050201}. This characterization is robust: if the marginals are perturbed, with each deviating from the ideal by a given trace-norm error, then the noisy stabilizer state deviates from the corresponding pure stabilizer state by at most the square root of the total marginal error~\cite{yu2026quantumstatedeterminabilitylocal}.
Furthermore, access to the reduced states enables efficient computation of local quantities such as the von Neumann entropy $S(\rho_A)$. As we can efficiently evaluate the entropy of small subsystems, we can use inequalities of the whole system entropy in terms of smaller system entropies \cite{Ruskai02}.  Hence, local Pauli expectations provide bounds to global mixedness and correlations.

\subsection{Bell inequality violations}
Bell-type inequalities consider a $n$-partite scenario in which party $i$ chooses one of two dichotomic observables, $A_{x_i}^{(i)}$ with $x_i \in \{0,1\}$, each producing outcomes \(\pm 1\). The resulting correlations are described by the correlators
\begin{equation}
\left\langle
A_{x_{i_1}}^{(i_1)}
\cdots
A_{x_{i_k}}^{(i_k)}
\right\rangle_{\rho}.
\end{equation}
In this setting, a general Bell-type inequality takes the form
\begin{equation}
\sum_{k=1}^{n}
\sum_{\substack{
1 \le i_1 \le \dots \le i_k \le n \\
x_1,\dots,x_n \in \{0,1\}
}}
\alpha_{x_1,\dots,x_n}^{i_1,\dots,i_k}
\,
\left\langle
A_{x_{i_1}}^{(i_1)}
\cdots
A_{x_{i_k}}^{(i_k)}
\right\rangle_{\rho}
\le \beta_C,
\end{equation}
where \(\beta_C\) denotes the classical bound. For graph states, such inequalities were first derived in Ref.~\cite{PhysRevLett.95.120405}. There are also Bell-type inequalities expressible as linear combinations of \(\mathcal{O}(n)\) multi-qubit Pauli correlators~\cite{PhysRevLett.124.020402}, which we can determine efficiently. Their violation quantifies nonlocal correlations beyond classical models. Since each term corresponds to a Pauli expectation value, the Bell score can be evaluated efficiently. This enables analytical studies of how noise degrades non-locality in stabilizer-state resources.

\subsection{Energy expectation values of Hamiltonians}
For Hamiltonians that can be written as a $\mathrm{poly}(n)$-term sum of Pauli operators,
\begin{equation}
    \langle H \rangle_\rho = \sum_i h_i \langle P_i \rangle,
\end{equation}
where $h_i \in \mathbbm{R}$, the energy can be computed directly from $\mathrm{poly}(n)$ Pauli expectation values. This assumption is not very restrictive. Any $n$-qubit Hamiltonian with only finite $k$-body interactions can be expanded in the Pauli basis as
\begin{equation}
H=\sum_i h_i P_i,
\end{equation}
where each $P_i$ is a Pauli string with support on at most $k$ qubits. For fixed $k$, the number of such terms scales as $\mathcal{O}(n^k)$, which is polynomial in $n$. Therefore, any Hamiltonian with finite $k$-body interactions admits a $\mathrm{poly}(n)$-term Pauli decomposition. This is particularly relevant for variational quantum algorithms with stabilizer state hot-starts~\cite{Sun2025stabilizerground} and benchmarking preparation of ground or low-energy states within the stabilizer formalism.

\subsection{Average MBQC fidelity.}

A recently introduced quality metric for noisy MBQC is the average MBQC fidelity~\cite{2603.13753}. Rather than evaluating the fidelity of the full resource state, one fixes the output vertices and averages over all measurement angles, yielding $\overline{F}_{\mathrm{MBQC}}(\rho)$. This quantity admits the compact form
\begin{equation}
    \overline{F}_{\mathrm{MBQC}}(\rho) = \mathrm{tr}(\rho\,\Omega),
\end{equation}
where $\Omega$ is a weighted sum of stabilizers. Hence, assessing the computational usefulness of a resource state reduces to calculating a subset of Pauli expectation values.

Efficient estimation is possible via sampling methods analogous to direct fidelity estimation~\cite{flammia2011direct}. In particular, sampling 
\[
m \geq \frac{2}{\epsilon^2}\log\!\left(\frac{2}{\delta^2}\right)
\]
stabilizer expectation values yields an $\epsilon$-accurate estimate of $\overline{F}_{\mathrm{MBQC}}(\rho)$ with success probability at least $1-\delta$~\cite{2603.13753}.

\section{Comparison of sampling-based and analytical expectation-value estimation}\label{app:sampling_vs_analytical_exp_values}

Here we compare the two conceptually distinct approaches to computing Pauli expectation values in the non-deterministic noisy stabilizer-circuit fragment.

The sampling-based approach generates measurement samples from the noisy circuit and estimates the desired expectation value from empirical frequencies. For a Pauli observable with outcomes in $\{\pm1\}$, the estimator is the sample mean over $M$ independent shots,
\begin{equation}
    \hat{\mu}=\frac{1}{M}\sum_{k=1}^M X_k,
    \qquad X_k\in\{\pm1\}.
\end{equation}
The standard Monte-Carlo error bar is the standard error of this estimator~\cite{Owen2013MonteCarlo}. Since
\begin{equation}
    \mathrm{Var}(X_k)=1-\mu^2,
    \qquad \mu=\mathbb{E}[X_k]=\langle P\rangle,
\end{equation}
the standard error of the sample mean is
\begin{equation}
    \mathrm{SE}(\hat{\mu})
    =
    \sqrt{\frac{1-\mu^2}{M}}
    \le \frac{1}{\sqrt{M}}.
\end{equation}
Hence, the statistical uncertainty decreases with the characteristic Monte-Carlo scaling
\begin{equation}
    \mathrm{SE}(\hat{\mu}) = O(M^{-1/2}).
\end{equation}
Equivalently, achieving a target additive accuracy $\epsilon$ requires
\begin{equation}
    M=O(\epsilon^{-2}).
\end{equation}

A corresponding rigorous finite-sample statement follows from Hoeffding's inequality~\cite{Hoeffding1963}. For $\{\pm1\}$-valued outcomes one has
\begin{equation}
    \Pr\!\left[\,|\hat{\mu}-\mu|\ge \epsilon\,\right]
    \le 2 e^{-M\epsilon^2/2},
\end{equation}
so it suffices to choose
\begin{equation}
    M \ge \frac{2}{\epsilon^2}\log\!\left(\frac{2}{\delta}\right)
\end{equation}
in order to guarantee additive error at most $\epsilon$ with probability at least $1-\delta$.

The practical advantage of this strategy is its low per-shot cost. After a single noiseless reference simulation, each additional shot only requires propagating a Pauli frame and updating the corresponding measurement parities, which is substantially cheaper than rerunning a full tableau simulation. Consequently, for modest target accuracy, sampling can be very efficient and may outperform an analytical evaluation.

The second approach is the analytical evaluation developed above. Rather than estimating expectation values from repeated circuit executions, one computes them directly from the Heisenberg action of the Pauli-diagonal noise channels on the observable. For a stabilizer element $S_l\in\mathcal{S}$, this yields
\begin{equation}
    \langle S_l\rangle_\rho
    =
    \prod_{i=1}^{N}
    \sum_j
    \lambda^{(i)}_j
    (-1)^{[[E_j^{(i)},S_l]]},
\end{equation}
while $\langle P\rangle_\rho = 0$ for $\pm P \notin \mathcal{S}$. The computational cost is therefore determined by evaluating the relevant commutation relations once, rather than by repeated Monte-Carlo sampling. In particular, for fixed noise structure, these commutation relations are independent of the numerical values of the noise strengths $\lambda_j^{(i)}$, so one obtains a parametric expression that can be reused efficiently for sweeps over noise parameters.

\section{Photonic graph state generation protocol} \label{app:graph_state_generation}
Here, we expand on the generation of the noisy graph-state resource states discussed in Sec.~\ref{subsec:example} and highlight connections to photonic platforms. First, caterpillar graph states are the only class of graph states, up to local Clifford operations, that can be generated by a single quantum emitter using local unitaries and measurements~\cite{PhysRevA.111.052604}. In photonic platforms, type-II fusion processes are probabilistic~\cite{PhysRevLett.95.010501}, so we focus on runs in which resource-state generation succeeds.

We now describe how a single emitter generates a photonic graph state. Producing a photonic caterpillar graph state requires entangled-photon emission, local Clifford operations, and local Pauli measurements. We model the emission process as the initialization of a photonic $\ket{+}$ state followed by a controlled-$Z$ gate between the emitter and the photon.

Next, we describe the specific caterpillar-generation protocol analyzed here. To do so, we introduce the graph-state manipulation known as local complementation, a local Clifford operation that graphically inverts the neighborhood of a target vertex. It is given by~\cite{PhysRevA.69.062311,Hein2006GraphStates}
\begin{equation}
    LC_{v}=(HSH)_v \bigotimes_{w \in N(v)} S_w,
\end{equation}
where $S$ denotes the Clifford phase gate.

\begin{figure}
    \centering
    \includegraphics[width=1\linewidth]{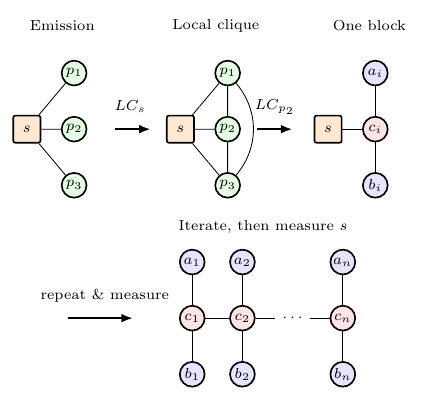}
    \caption{Illustration of a generation protocol of a photonic caterpillar graph state, by a single quantum emitter.}
    \label{fig:7}
\end{figure}

A two-spike caterpillar can be generated from a single emitter by repeatedly emitting three photons and then applying two local complementations. After the three emissions, the emitter is connected to the three new photons in a local star configuration. A local complementation at the emitter turns the three emitted photons into a clique. A second local complementation at one of these photons transforms the block into one corpus node connected to two spike nodes, while keeping the emitter attached to the corpus so that the procedure can be repeated. Note that the two local complementations can be combined into two local Clifford gates:
\begin{equation}
    LC_{p}LC_s= (HSHS)_{p} (SHSH)_s
\end{equation}
Iterating this three-photon block builds the full caterpillar backbone with two spikes per corpus node. A final $Z$-basis measurement of the emitter removes it, leaving the desired photonic caterpillar graph state; see Fig.~\ref{fig:7} for an illustration.

\section{Time complexity of circuit compression}\label{app:compression_complexity}

We represent each Pauli operator by a $2n$ binary symplectic vector and a constant-size phase label, and maintain a full stabilizer-destabilizer tableau of size $\mathcal{O}(n^2)$.
A Pauli multiplication or commutation test takes $\mathcal{O}(n)$ elementary binary operations~\cite{PhysRevA.70.052328}.
The initial tableau is supplied as input or initialized for a standard product state in $\mathcal{O}(n^2)$ time.
If only stabilizer generators are supplied, constructing compatible destabilizers by Gaussian elimination adds an $\mathcal{O}(n^3)$ preprocessing cost.

At a given update, let $m$ be the number of retained measurement projectors, $r\le m$ the number classified as random, and $k$ the number of stored Pauli noise terms.
Then $m\le M$, $r\le M$, and $k\le K$.
Each conjugation or stabilizer adjustment maps a Pauli term to a single Pauli term, so the rules do not increase the number of terms in a channel. The costs below count operations on Pauli labels, while noise weights are carried along unchanged.

\paragraph{Noise channels.}
Appending a channel with $K_i$ explicitly listed Pauli terms costs $\mathcal{O}(nK_i)$ for reading and storing its Pauli labels.
Summed over all channels, this contributes $\mathcal{O}(nK)$.

\paragraph{Deterministic measurements.}
Classifying an arbitrary Pauli observable and determining its sign relative to the reference state takes at most $\mathcal{O}(n^2)$ tableau operations~\cite{PhysRevA.70.052328}.
Scanning the retained projectors and noise terms to test whether the new projector can be propagated or absorbed costs at most $\mathcal{O}(n(m+k))$. Thus,
\begin{equation}
    T_D(m,k)=\mathcal{O}\!\left(n^2+n(m+k)\right).
\end{equation}

\paragraph{Clifford gates.}
An elementary one- or two-qubit Clifford gate modifies only a constant number of entries in each Pauli vector.
Conjugating all retained measurement observables and noise terms therefore costs $\mathcal{O}(m+k)$, while updating the reference tableau costs $\mathcal{O}(n)$~\cite{PhysRevA.70.052328}.
The total cost per gate is
\begin{equation}
    T_C(m,k)=\mathcal{O}(n+m+k).
\end{equation}
An arbitrary multi-qubit Clifford is counted through its decomposition into elementary gates.

\paragraph{Random measurements.}
First, checking commutation with the retained measurement observables costs $\mathcal{O}(nm)$.
To test Eq.~\eqref{equ:cond_stabilizer_existence}, form the binary matrix whose rows are $\boldsymbol{\alpha}^0,\ldots,\boldsymbol{\alpha}^r$.
The coefficients can be obtained directly as $\alpha_i^j=[[g_i,P_j^{(R)}]]$, so constructing all $r+1$ rows costs $\mathcal{O}(n^2(r+1))$.
Gaussian elimination on the resulting $(r+1)\times n$ system, also takes at most $\mathcal{O}(n^2(r+1))$ operations.
This bound allows the system to be rebuilt at each measurement.

If the system admits a solution, constructing the adjustment stabilizer from at most $n$ generators costs $\mathcal{O}(n^2)$.
Checking and, where necessary, multiplying the stored noise terms by this stabilizer costs $\mathcal{O}(nk)$, and updating the reference tableau costs $\mathcal{O}(n^2)$.
If absorption fails, only the new projector is appended after the required checks.
In either case, the per-measurement bound is
\begin{equation}
    T_R(m,r,k)
    =\mathcal{O}\!\left(n^2(r+1)+n(m+k)\right).
\end{equation}

\paragraph{Total cost.}
Summing over $C$ Clifford gates, $M-R$ deterministic measurements, and $R$ random measurements, and using $m\le M$, $r\le M$, and $k\le K$, gives Eq.~\eqref{eq:compression_complexity}, including initialization and noise-input storage. The bound is polynomial in $n$, $C$, $M$, and the explicit noise-term count $K$.

\section{Special non-diagonal noise channels} \label{app:special_non_diag_channels}
Here, we discuss special classes of non-diagonal noise channels and mixed input states, where the noise-channel updates are independent of the measurement outcome of the non-deterministic measurements. 
\subsection{Stochastic mixture of stabilizer states} \label{stoch:mixture of stabilizer states}
One can consider simulating a stochastic mixture of stabilizer states
\begin{equation}
    \rho=\sum_{i} c_i \ketbra{S^{(i)}},
\end{equation}
 where $\ketbra{S^{(i)}}$ are different stabilizer states. 
 Considering the noisy simulation of such diagonal states
\begin{equation}
    \rho=\sum_{i} c_i  \mathcal{E}^{i}_1 \cdots \mathcal{E}^{i}_n \ketbra{S^{(i)}},
\end{equation}
we differentiate again between Clifford update rules, and Pauli-measurement update rules. For the Clifford updates, we treat every superposition term independently and update the associated noise channel string and the state by the standard NSF rules. 

For the non-deterministic Pauli measurement, i.e. the measurement is non-deterministic for all stabilizer states $\ketbra{S^{(i)}}$, we also treat each term individually and using the standard NSF rules we can update the terms independently, and compute the normalization, i.e. outcome probability, similar to Sec.~\ref{sec:gen-rot}.
Note, in contrast to scheme before, see Sec.~\ref{sec:gen-rot}, where we kept, one stabilizer state representative fixed, and modified it by potentially off diagonal channels, we do not have outcome dependent noise channel update rules, as the Pauli measurement correction enter as a Pauli unitary in each superposition term:
\begin{equation}
\begin{aligned}
&P_{Z,\pm}\mathcal{E}^{i}_1\cdots\mathcal{E}^{i}_n \ketbra{S^{(i)}} \\
&=\tilde{\mathcal{E}^{i}}_1\cdots\tilde{\mathcal{E}^{i}}_n J^{(i)}(\pm)\ketbra{\tilde{S}^{(i)}}J^{(i)}(\pm),
\end{aligned}
\end{equation}
where $J^{(i)}(\pm)$ is the Pauli measurement correction which changes the stabilizer state, dependent on the outcome. Hence, for the next updates the updated noise channels will be independent of the measurement outcome, as the phase cancels. Furthermore, the outcome probability is uniform for each measurement branch, and the coefficients need not to be differently updated, dependent on the measurement outcome. 

\subsection{Clifford and Pauli measurement noise channels}
One can also consider the incorporation of noise channels, which are probabilistic applications of operations, which are efficiently treatable by the standard NSF framework. Specifically, one can consider the probabilistic application of Clifford unitaries as well as non-deterministic Pauli measurements. We denote, the joint group of Pauli measurements and Clifford unitaries as $CM$, and denote error channels diagonal in this group $\mathcal{E}_{CM}$. These, error channels were studied in the context of approximating non-Pauli-diagonal error channels \cite{PhysRevA.87.030302, PhysRevA.87.012324}, and were shown to approximate important channels, such as the Amplitude damping channel, better than Pauli-diagonal channels. Hence, given a state given by the standard representation of the NSF $\mathcal{E}_1 \cdots \mathcal{E}_N \ketbra{S}$ and we apply a channel $\mathcal{E}_{CM}$, then we multiply out the channel $\mathcal{E}_{CM} (\bullet) = \sum_k \lambda_k Q_k (\bullet) Q_k$, with $Q_k \in CM$, and for each term we utilize the standard NSF, yielding
\begin{equation}
    \mathcal{E}_{CM} \mathcal{E}_1 \cdots \mathcal{E}_N \ketbra{S}=\sum_{k} \lambda_k \tilde{\mathcal{E}}^{(k)}_1 \cdots \tilde{\mathcal{E}}^{(k)}_N \ketbra{S^{(k)}},
\end{equation}
where $\ket{S^{(k)}}=Q_k \ket{S}$. Hence, we are in the same situation as in Appendix~\ref{stoch:mixture of stabilizer states}, and can recast the methods from there.

\end{document}